\documentclass[%
 reprint,
superscriptaddress,
 amsmath,amssymb,
 aps,
]{revtex4-2}

\usepackage{graphicx}
\usepackage{dcolumn}
\usepackage{bm}
\usepackage[caption=false]{subfig}
\usepackage{hyperref}
\usepackage[percent]{overpic}
\usepackage{cleveref}
\usepackage{physics}
\usepackage{orcidlink}
\usepackage{placeins}
\begin{document}

\title{Trapping and cooling mechanisms in blue-detuned magneto-optical traps of molecules}

\author{Qinshu~Lyu\orcidlink{0009-0007-6310-5714}}
\affiliation{Centre for Cold Matter, Blackett Laboratory, Imperial College London, London SW7 2AZ, United Kingdom}
\affiliation{Department of Physics, Harvard University, Cambridge, MA 02138, USA}
\affiliation{Harvard-MIT Center for Ultracold Atoms, Cambridge, MA 02138, USA}

\author{M.~R.~Tarbutt\orcidlink{0000-0003-2713-9531}}
\email{m.tarbutt@imperial.ac.uk}
\affiliation{Centre for Cold Matter, Blackett Laboratory, Imperial College London, London SW7 2AZ, United Kingdom}

\begin{abstract}
In red-detuned magneto-optical traps (MOTs) of molecules, sub-Doppler heating competes with Doppler cooling, resulting in high temperature and low density. A solution is offered by the blue-detuned MOT where sub-Doppler cooling dominates and the cloud is compressed. Several blue-detuned molecular MOTs have been implemented. A recent implementation relies on a pair of orthogonally polarized components whose frequency separation is smaller than the transition linewidth. We identify the trapping force in these MOTs. At a certain magnetic field, there is a state that is dark to the laser propagating in one direction, but not to the counter-propagating one. This Zeeman-induced dark state (ZIDS) sets up an imbalance in the photon scattering rate, leading to a restoring force. We also study the role of the moving lattices generated by the closely-spaced frequency components of the light. We show that there is a velocity-dependent force that drives the molecules towards the speeds of these moving lattices, and that over a relevant range of magnetic fields this combines with the ZIDS force to transport molecules towards the centre of the MOT. Here, gray molasses cooling, assisted by non-adiabatic transitions driven by the time-varying polarization of the light field, cools the molecules towards zero velocity. We study these mechanisms for model systems with simple level structures, then extend them to molecules with ground state hyperfine structure.

\end{abstract}

\maketitle

\section{Introduction}

The past decade has seen the rapid development of direct laser cooling and trapping of molecules, both diatomic~\cite{Barry2014, Truppe2017b, Collopy2018, Zeng2024, Padilla-Castillo2025, Fitch2021b} and polyatomic~ \cite{Vilas2022, Lasner2025, Augenbraun2023}. These advances have been driven by applications in many-body physics~\cite{Cornish2024}, quantum computation~\cite{Bao2023, Holland2023, Zhang2022b}, ultracold chemistry~\cite{Karman2024} and tests of fundamental physics~\cite{Fitch2020b, Anderegg2023, Athanasakis-Kaklamanakis2025, DeMille2024}. Experiments with trapped, laser-cooled molecules typically start with a magneto-optical trap using light red-detuned from a type-II transition ($F\geq F'$ where $F,F'$ are the angular momenta of the ground and excited states). The type-II transition is needed for rotational closure, and the negative detuning is needed for Doppler cooling. However, this configuration leads to sub-Doppler heating~\cite{Devlin2016}, resulting in temperatures above the Doppler limit, and large, dilute clouds. A subsequent phase of blue-detuned optical molasses uses sub-Doppler cooling to reduce the temperature to a few $\mu$K~\cite{Truppe2017b, Cheuk2018, Caldwell2019}, but there is no restoring force in the molasses so the cloud remains at low density. This greatly reduces the loading efficiency into optical dipole traps, and the low initial phase space density inhibits effective evaporative cooling into the quantum degenerate regime. 

The blue-detuned MOT, first proposed in \cite{Devlin2016}, solves these problems by providing confining forces and sub-Doppler cooling simultaneously. The method was first demonstrated for Rb atoms~\cite{Jarvis2018} and shown to produce MOTs of high density and low temperature. Subsequently, the technique has been applied to great effect for molecules~\cite{Burau2023, Li2024, Jorapur2024, Yu2026, Hallas2024arxiv, Zeng2025arxiv}, resulting in molecular MOTs with densities exceeding $10^{10}$~cm$^{-3}$ that can load optical dipole traps at phase-densities exceeding $10^{-6}$. Various blue-detuned MOT schemes have used for various molecular species. Here, we focus on a scheme applied to CaF~\cite{Yu2026}, CaOH~\cite{Hallas2024arxiv} and BaF~\cite{Zeng2025arxiv} where two closely-spaced frequency components of opposite handedness generate a pair of moving lattices travelling in opposite directions. These moving lattices can transport molecules towards the centre of the MOT, so the scheme has been called a `conveyor belt' MOT. While the conveyor mechanism is described in \cite{Li2025}, the processes involved have not been fully understood and the mechanism does not fully explain the trapping and cooling present in this MOT. 

In this paper, we identify the mechanism that produces confinement around the magnetic field zero. At a specific non-zero magnetic field, whose value is controlled by the spacing of the two frequency components of the light, the molecules become dark to one of the two counter-propagating beams, resulting in preferential photon scattering from the beam that pushes the molecules towards the field zero. We refer to this as the Zeeman-induced dark state (ZIDS) trapping mechanism. We also study the conveyor belt mechanism. A moving molecule is dragged towards the speed of the conveyor belt moving in the same direction as the molecule, i.e. it is driven into whichever conveyor is moving more slowly in the molecule frame. At large magnetic fields, both conveyors are active and molecules may be transported in either direction, introducing a loss mechanism into the MOT. At intermediate fields however, the velocity-dependent force due to the moving lattices combines with the ZIDS mechanism to drive molecules preferentially into the lattice that transports them towards the MOT centre. For small magnetic fields, the molecules are cooled towards zero velocity by gray molasses cooling, but with nonadiabatic transitions from dark to bright states driven by the time-varying polarization of the light field as well as motion through the light field.

In Sec.~\ref{sec:model_1D}, we study theses forces in one dimension, for a model system that has $F=F'=1$. In Sec.~\ref{sec:model_3D}, we show that the same mechanisms operate in three dimensions. Then, in Sec.~\ref{sec:model_two_hyperfine}, we extend our study to the case of a molecule with ground-state hyperfine structure.

\section{Model in one dimension}\label{sec:model_1D}

To help understand the trapping and cooling mechanisms, we start with the simplest of schemes, a two-level molecule with angular momentum $F=1$ in the ground state and $F'=1$ in the excited state, and decay rate $\Gamma$. The molecule interacts with counter-propagating laser beams in one dimension only. In the standard 1D $\sigma^{+}\sigma^{-}$ MOT configuration, with only a single frequency in each beam, the force is zero at all positions and velocities~\cite{Chang2002, Tarbutt2015, Devlin2016}. By contrast, when there are two closely-spaced frequency components in each beam, we find that there is a velocity-dependent force whose sign reverses with $\Delta$, and a position-dependent force whose sign reverses with $\delta$. Our aim is to understand how these forces arise.

\begin{figure}
    \subfloat[]{
        \includegraphics[width=0.45\linewidth]{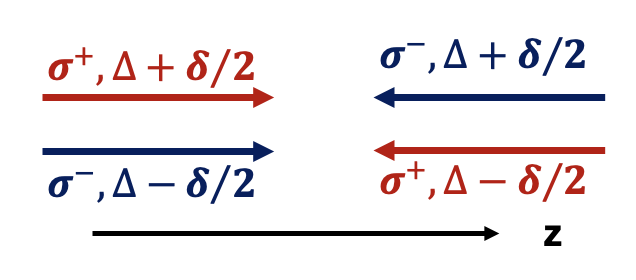}
        \label{fig:laser_config}
        }
    \hfill
    \subfloat[]{
        \includegraphics[width=0.45\linewidth]{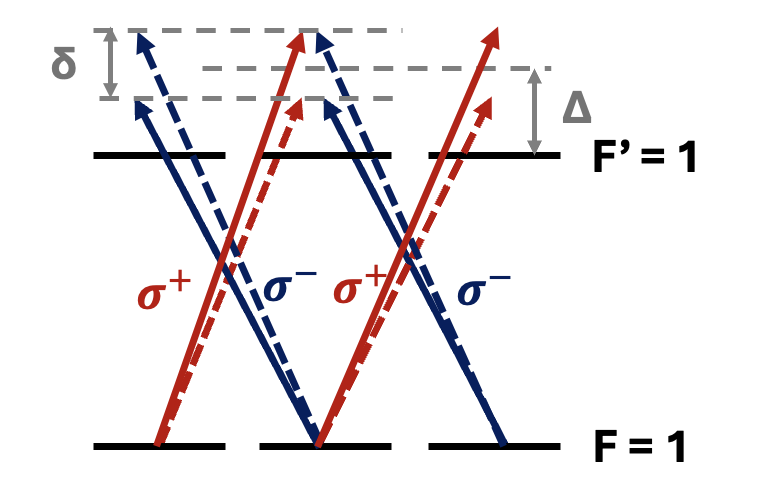}
        \label{fig:mot_scheme}
        }
       \caption{A simple scheme for a blue-detuned MOT in 1D. (a) Laser configuration, showing polarizations and detunings. The laser components have a global detuning of $\Delta$ and a frequency difference of $\delta$. The polarizations are denoted $\sigma^{\pm}$ according to whether the light drives $\Delta m=\pm 1$ transitions, where $m$ is the projection of the angular momentum onto the $z$-axis. (b) The $F=1 \rightarrow F'=1$ system, showing transitions driven by the lasers. Solid (dashed) arrows represent lasers propagating towards $+z(-z)$.}
    \label{fig:scheme_1_1}
\end{figure}

Figure \ref{fig:scheme_1_1} illustrates the scheme. Each laser beam contains two components with opposite circular handedness, with mean frequency $\omega$, difference frequency $\delta$, amplitude $E_0$ and wavelength $\lambda=2\pi/k$. Their average detuning from the molecular transition frequency is $\Delta$, and the intensity of each beam, in units of the saturation intensity, is denoted $s$. In the beam directed towards $\pm z$, the $\sigma^{\pm}$ component has the higher frequency. The total electric field of the light is
\begin{align}
    \vec{E}&=\hat{x}\, 4E_0 \cos(k z)\cos(\delta t/2)\cos(\omega t) \nonumber\\ &+ \hat{y} \,4E_0\sin(k z)\sin(\delta t/2)\sin(\omega t) \\
    &= -\hat{e}_{+1}\,\sqrt{2}E_0 \cos(k z+\delta t/2)e^{i\omega t} \nonumber \\&+\hat{e}_{-1}\,\sqrt{2}E_0\cos(-k z+\delta t/2)e^{i\omega t}+c.c.\,,
    \label{eq:lightField}
\end{align}
where $\hat{e}_{\pm 1}=\mp\frac{1}{\sqrt{2}}(\hat{x}\pm i \hat{y})$ are the unit circular polarization vectors.
The first form may be interpreted as a pair of standing waves oscillating with frequency $\delta/2$, one polarized along $\hat{x}$, the other along $\hat{y}$, and $\pi/2$ out of phase in both space and time. At $\delta t = 0$ ($\delta t = \pi$), the light is polarized along $\hat{x}$ ($\hat{y}$) everywhere, and the standing wave has maximum contrast. In between, the polarization oscillates in space and the standing wave has a lower contrast. The second form may be interpreted as a pair of counter-propagating lattices of opposite circular handedness moving with speeds $\pm \delta/(2k)$.

\subsection{Simulation results}

We numerically solve the optical Bloch equations (OBEs) for this system using the methods described in \cite{Devlin2016} and implemented in \cite{Eckel2022}, and thus determine the force as a function of velocity, $v$, and magnetic field, $B$ (equivalent to position, $z$, in the MOT). We average the force over a time period of either $10\lambda/v$ or $200\pi/\Gamma$, whichever is smallest, and continue to integrate the OBEs until the force converges. We repeat this $N_{\rm rep}$ times (typically $N_{\rm rep}=256$), choosing the initial position at random each time from an interval of size $\lambda$. The relative phase of the co-propagating beams is a constant, whereas the phase difference of the counter-propagating beams is chosen at random on each iteration. We find the mean and standard deviation of the set of $N_{\rm rep}$ results. We then repeat this procedure for many different values of $v$ and $B$ to build a map of the force.

\begin{figure}
    \subfloat[$\Delta=2\Gamma$, $\delta=0.2\Gamma$]{
        \includegraphics[width=0.48\linewidth]{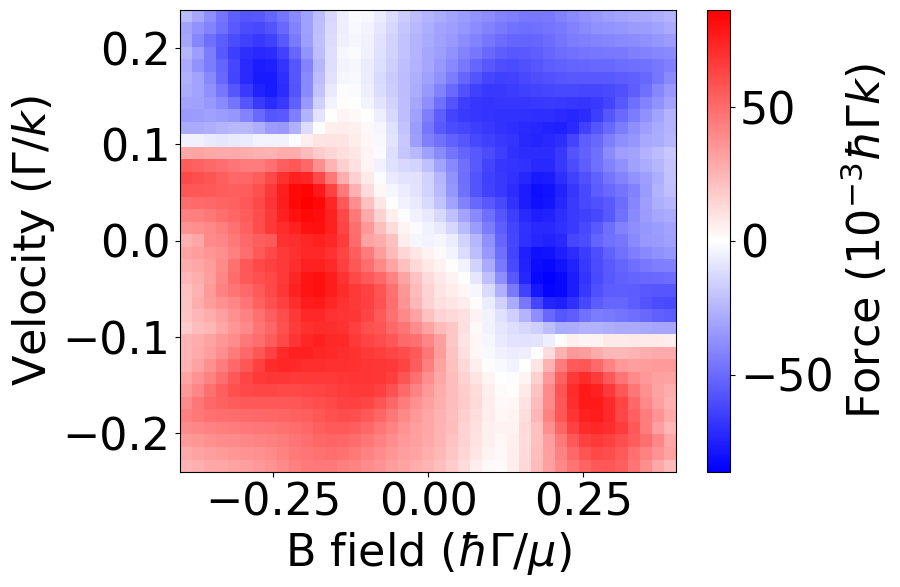}
        \label{fig:heatmap_D_2_d_p2}
        }
    \subfloat[$\Delta=2\Gamma$, $\delta=-0.2\Gamma$]{
        \includegraphics[width=0.48\linewidth]{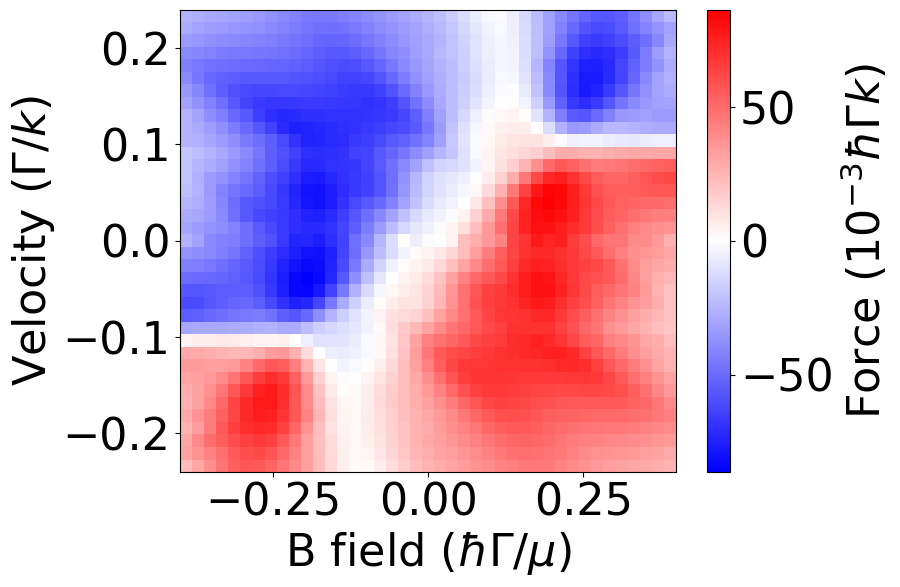}
        \label{}
        }
    \hfill
    \subfloat[$\Delta=-2\Gamma$, $\delta=0.2\Gamma$]{
        \includegraphics[width=0.48\linewidth]{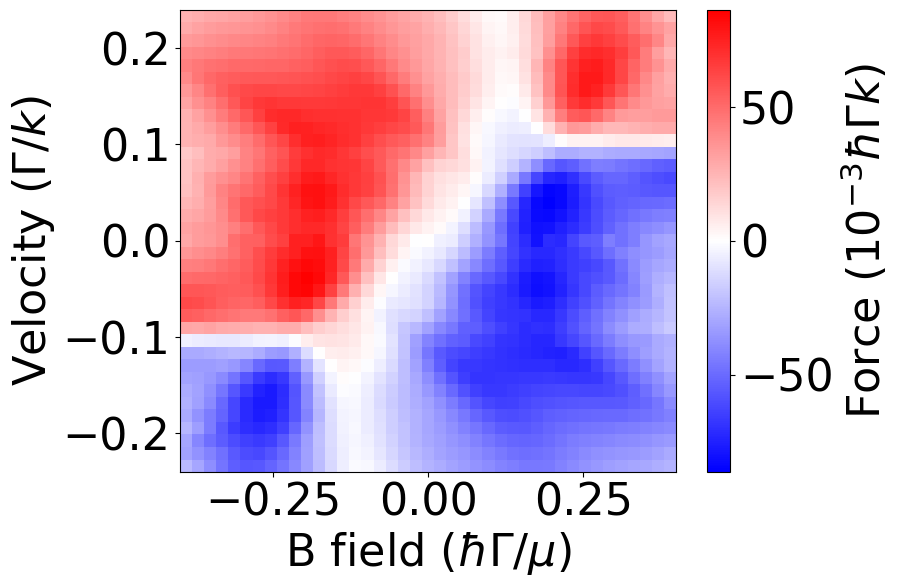}
        \label{fig:heatmap_D_m2_d_p2}
        }
    \subfloat[$\Delta=-2\Gamma$, $\delta=-0.2\Gamma$]{
        \includegraphics[width=0.48\linewidth]{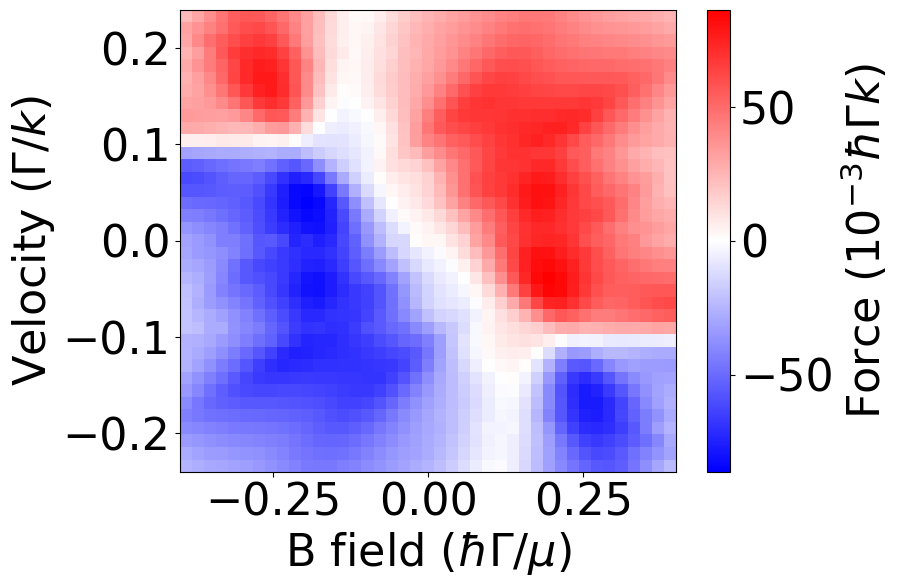}
        \label{fig:heatmap_D_m2_d_mp2}
        }
    \caption{Force as a function of $v$ and $B$ for the 1D system illustrated in Fig.~\ref{fig:scheme_1_1}, determined from OBE simulations. We have used $s=5$ per beam and $N_{\rm rep}=256$. }
    \label{fig:1D_heatmap}
\end{figure}

Figure \ref{fig:1D_heatmap} shows the results of these simulations. Throughout our discussion, we assume the MOT has a positive magnetic field gradient, $dB/dz>0$. At zero velocity, molecules are trapped around $B=0$ when $\delta>0$ (force is negative at positive $z$) and anti-trapped when $\delta<0$. The sign of $\Delta$ does not affect the trapping. Close to $B=0$ the molecules are cooled when $\Delta>0$ (force is negative at positive $v$), and are heated when $\Delta<0$. This is characteristic of sub-Doppler cooling in type-II systems. For these small values of $B$, the sign of $\delta$ does not affect the cooling. At larger values of $B$, and for positive $\Delta$, the molecules are instead cooled to the velocity of one of the two counter-propagating lattices. Which of the two is determined by the sign of $B$ -- at positive $B$ the molecules are driven to speed $-\delta/(2k)$. At negative $\Delta$, molecules are heated away from these special velocities. For intermediate values of $B$ (and positive $\Delta$) the molecules are cooled towards a velocity between zero and $\pm \delta/(2k)$.

In the following, we aim to understand the features shown in Fig.~\ref{fig:1D_heatmap}. What is the trapping mechanism and why does it reverse with $\delta$? What is the mechanism that cools molecules to zero velocity? What is the mechanism that drives the molecules to the lattice velocities and why does the sign of $B$ determine which of the two lattices is preferred?

\subsection{Trapping mechanism\label{sec:trapping_mechanism}}

\begin{figure}
    \subfloat[]{
        \includegraphics[width=0.4\linewidth]{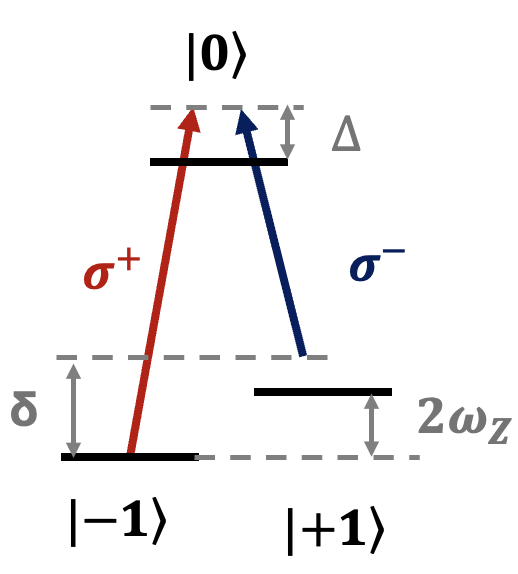}
        \label{fig:pz_lambda}
        }
    \subfloat[]{
        \includegraphics[width=0.4\linewidth]{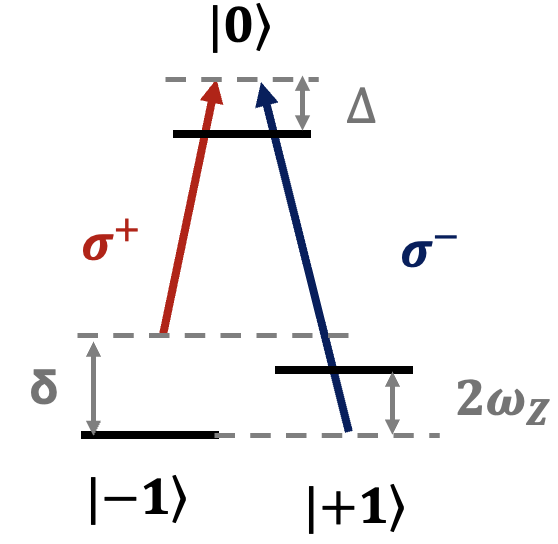}
        \label{fig:mz_lambda}
        }
    \hfill
    \subfloat[]{
        \includegraphics[width=0.45\linewidth]{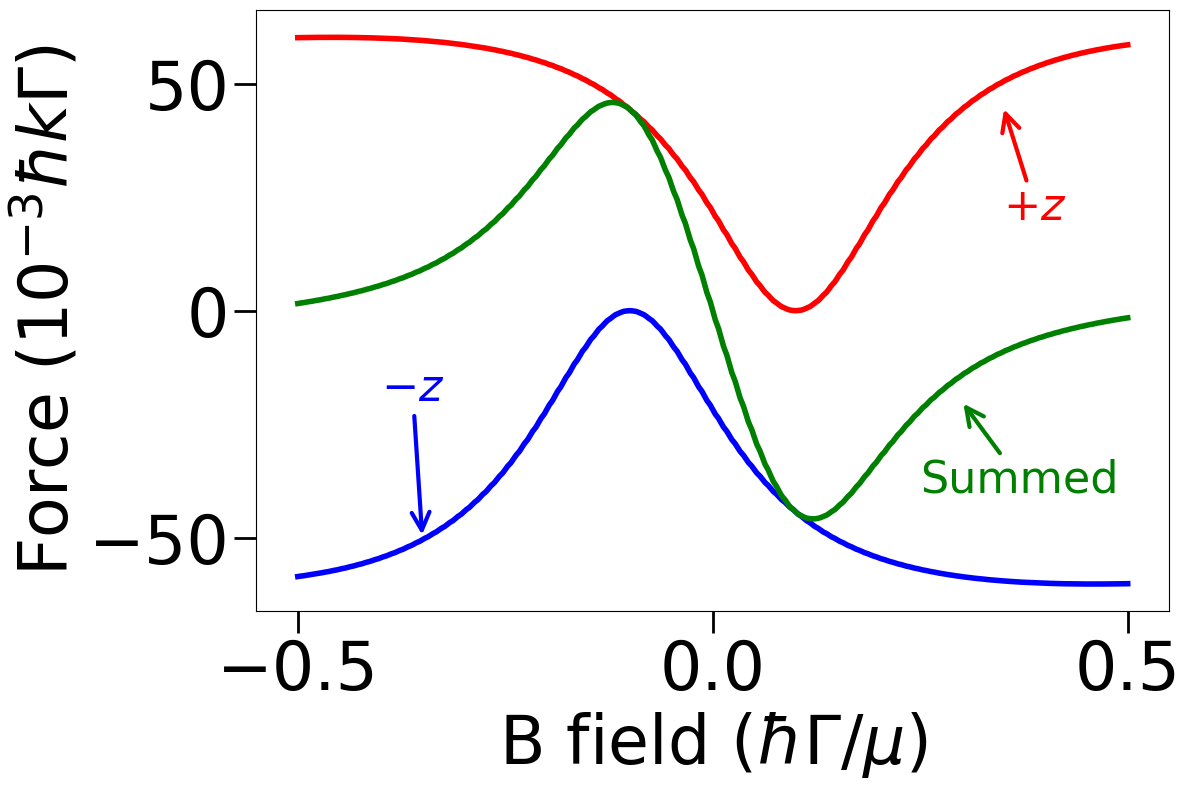}
        \label{fig:coprop_force}
        }
    \subfloat[]{
        \includegraphics[width=0.45\linewidth]{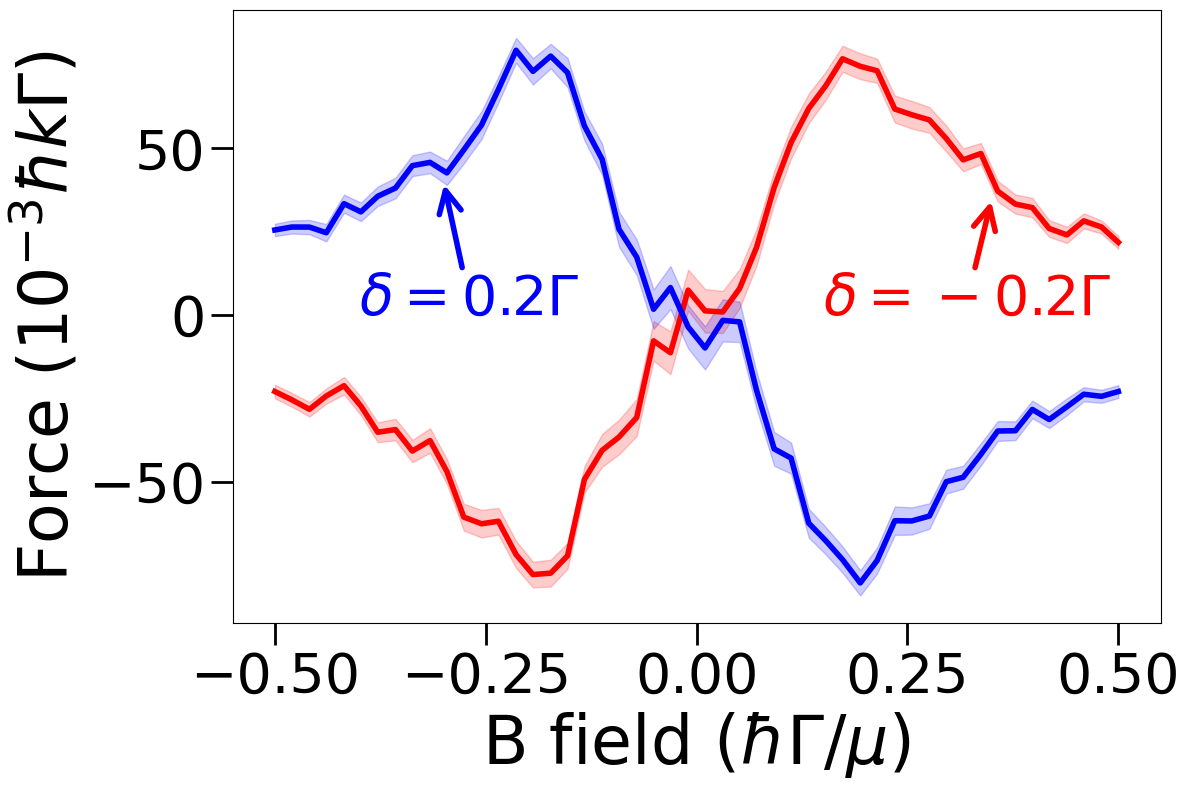}
        \label{fig:1D_force}
    }    
    \caption{$\Lambda$ system formed by (a) $+z$ laser pair and (b) $-z$ laser pair at $B>0$. The magnetic field brings the two photon detuning towards zero for the $+z$ laser pair, introducing a stable dark state and lowering the scattering rate from the $+z$ laser. The field increases the two-photon detuning for the $-z$ laser pair, so the molecule scatters more photons from this direction. $\omega_Z=\mu B/\hbar$ is the Zeeman shift. (c) Force curve for a pair of co-propagating lasers in the $+z$ (red) and $-z$ (blue) directions. Reduced scattering rate near the dark state resonance can be observed. The sum of the forces (green) leads to trapping. (d) Force curve in 1D with both $\pm z$ laser pairs present, as in Fig.~\ref{fig:laser_config}. $\delta>0$ gives trapping and $\delta<0$ gives anti-trapping.}
    \label{}
\end{figure}

Let the magnetic moment be $\mu$ in the ground state and zero in the excited state~\footnote{This is a good model for many molecules cooled on the transition $A^2\Pi_{1/2}\leftarrow X^2\Sigma$.}, and let the spontaneous emission rate be $\Gamma$. First consider the case where only the beam pointing in the $+z$ direction is present, yielding the system shown in Fig.~\ref{fig:pz_lambda}. The two laser components form a $\Lambda$ system  with a dark state $\ket{d_+}=\frac{1}{\sqrt{2}}(e^{i(2\omega_{\rm Z} - \delta)t}\ket{-1}+\ket{+1})$, where $\pm\omega_{\rm Z}=\pm\mu B/\hbar$ is the Zeeman shift of the $m=\pm1$ levels. At the critical magnetic field $B_{\rm c}=\hbar\delta/(2\mu)$, the Zeeman shift matches $\delta$, the two-photon detuning goes to zero, $\ket{d_+}$ becomes time-independent and there will be no photon scattering. The force due to this beam goes to zero at $B_{\rm c}$. Now introduce the beam pointing in the $-z$ direction. The state that is dark to this beam is $\ket{d_-}=\frac{1}{\sqrt{2}}(e^{i(2\omega_{\rm Z} + \delta)t}\ket{-1}+\ket{+1})$, as can be seen in Fig.~\ref{fig:mz_lambda}. At $B_{\rm c}$, this state rotates at $2\delta$, rapidly becoming bright. The excited state has $m=0$ which can decay to the dark state or the orthogonal bright state with equal probability. When it decays to the bright state the atom can be excited by either laser beam, but when it decays to the dark state it can only be excited by the $-z$ beam. This sets up an imbalance in radiation pressure. The same arguments apply at $B=-B_{\rm c}$ but with everything reversed -- here the molecule preferentially scatters photons from the $+z$ beam. Thus, when $\delta>0$ and $dB/dz >0$ there will be a restoring force due to the difference in scattering rates set up by the Zeeman-induced dark states (ZIDS).

Figure \ref{fig:coprop_force} shows the forces from the $+z$ beams alone, and from the $-z$ beams alone. We have set $\Delta=2\Gamma$, $\delta = 0.2\Gamma$ and $s=5$. As expected from the discussion above, the force due to the light in the $+z$ ($-z$) direction goes to zero when $B=B_{\rm c}$ ($B=-B_{\rm c}$). Also shown in the figure is the sum of the two forces. While we should not expect this to be accurate, since the state that is dark to one beam is destabilized by the other, the summed force gives an intuitive picture of what to expect. Next, we simulate the full configuration shown in Fig.~\ref{fig:laser_config} with all four components present. Figure \ref{fig:1D_force} shows the force obtained in this case. We see that the true force is similar to the sum shown in Fig.~\ref{fig:coprop_force}, and that the force reverses when $\delta$ reverses, so we conclude that the ZIDS mechanism is responsible for the trapping force in this blue-detuned MOT. 

It is interesting to compare this ZIDS force with the na\"ive MOT force obtained by ignoring any dark states and just summing the scattering forces due to each component of the light, using the standard expression for the scattering force and accounting for the detuning of each component and the Zeeman shift of the transition that each drives. Interestingly, this also gives a position-dependent force whose sign depends on $\delta$. For $\delta>0$,  $dB/dz >0$, it provides a trapping force because the beam pushing the molecules back towards the centre is Zeeman-shifted closer to resonance. This force acts over a broad range of magnetic fields (determined by $\Delta$), but near the trap centre is far weaker than the force set up by the Zeeman-induced dark states. 

\subsection{Cooling mechanism at zero magnetic field\label{sec:cooling_zero_B}}

We turn now to the velocity-dependent part of the force. Recall that for the 1-1 system in 1D, with just a single polarization component in each beam, there is no force because the molecule decays with equal probability to $m=\pm 1$ so must scatter an equal number of photons from the $\sigma^{\pm}$ beams. For our configuration (Fig.~\ref{fig:laser_config}), this constraint is removed because both polarizations are present in both beams. For positive $\Delta$, there will be Doppler heating at large velocity and sub-Doppler cooling at low velocity. The sub-Doppler cooling is similar to the lin-$\phi$-lin, or gray molasses, mechanism~\cite{Devlin2016}. In the light field described by Eq.~(\ref{eq:lightField}), and at zero $B$, there is a dark state
\begin{equation}
    \ket{d}=\cos(k z-\delta t/2)\ket{-1} + \cos(k z+\delta t/2)\ket{+1}, 
    \label{eq:dark_superposition}
\end{equation}
where we have omitted the normalization. The dark state has zero energy while the orthogonal bright state has energy
\begin{align}
    E_{\rm b} &= \frac{\hbar \Delta}{2}\left(\sqrt{1+2\frac{\Omega^2}{\Delta^2}(1+\cos(2 k z)\cos(\delta t))}-1\right) \nonumber \\
    &\approx \hbar\zeta\left(1+\cos(2 k z)\cos(\delta t)\right),
    \label{eq:Eb}
\end{align}
where $\Omega$ is the Rabi frequency, $\zeta =\frac{\Omega^2}{2\Delta}$, and the approximation holds for $\Delta \gg \Omega$.
The bright state energy is proportional to the intensity of the light and goes to zero at the special points in time and space $(t,z)=(2n\pi/\delta, (2m+1)\lambda/4)$ and $(t,z)=((2n+1)\pi/\delta, m\lambda/2)$ where $n,m$ are integers. Optical pumping into the dark state tends to happen when the intensity is high, which is also where $E_{\rm b}$ is large, i.e. near the top of the hill. Transitions back to the bright state occur near the special points where $E_{\rm b}$ is close to zero, i.e. near the bottom of the hill. These transitions can either be driven by motion through the light field, or by the time-variation of the light field.

To understand the non-adiabatic transitions between dark and bright states, it is helpful to use a Landau-Zener treatment. In the basis of the two ground $m$-sublevels $\ket{\pm 1}$, a suitable effective Hamiltonian is
\begin{equation}
    H_{\rm LZ} = \Lambda I + V \sigma_x + \epsilon \sigma_z,
    \label{eq:HLZ}
\end{equation}
where $I$ is the identity operator, $\sigma_{x,z}$ are the Pauli operators and
\begin{align}
    \Lambda &= \frac{\hbar\zeta}{2}\left( 1+\cos(2kz)\cos(\delta t)\right),\\
    V &= \frac{\hbar\zeta}{2}\left( \cos(2 k z)+ \cos(\delta t)\right),\\
    \epsilon &= -\frac{\hbar\zeta}{2}\sin(2 k z)\sin(\delta t).    
\end{align}
The eigenstates of $H_{\rm LZ}$ are the dark and bright states and the eigenvalues are their energies (in the approximate form of Eq.~(\ref{eq:Eb})). Equation (\ref{eq:HLZ}) is in a standard form for a Landau-Zener treatment with an avoided crossing where $\epsilon=0$ and an energy gap of $2V$. To apply this treatment, we will approximate $\epsilon$ as varying linearly in time and approximate $V$ to be constant close to the crossings. 

First consider diabatic crossings driven by motion through the light field. Let the molecule be at $k z=\pi/2$ at time $t_0$ so that we can write $2k z= \pi+2k v(t-t_0) $. Linearizing around this point, we have $\epsilon \approx \hbar \zeta k v\sin(\delta t_0)(t-t_0)$ and $V\approx -\hbar\zeta\sin^2(\delta t_0/2)$. Applying the standard Landau-Zener result we find that the probability of a diabatic transition at this crossing is
\begin{equation}
    P_{\rm LZ}^{(1)} \approx \exp\left(-\pi \frac{\zeta}{k|v|}\frac{\sin^4(\delta t_0/2)}{|\sin(\delta t_0)|} \right).
    \label{eq:PLZ1}
\end{equation}
A similar result is obtained for all points where $2k z=n\pi$. We see that the probability is controlled by the dimensionless parameter $\zeta/(kv)$ which is the ratio of the light shift to the Doppler shift. When $\zeta/(kv) \gg 1$ (low $v$), the diabatic transitions can only occur if $\delta t_0 \approx 2m\pi$, so molecules have to pass through the right points at the right times for the cooling to be effective. When $\zeta/(kv) \ll 1$ (high $v$) there can be diabatic transitions for almost any value of $t_0$, the dark state is unstable, and cooling will be ineffective. The cooling will be most effective for intermediate values of $\zeta/(kv)$. We are typically interested in parameters where $\zeta \sim \Gamma$ and $k v<0.1$, giving $\zeta/(kv)>10$. This suggests that transitions driven by the motion will be quite strongly localized at the bottom of the potential hills. At very low velocity, the localization is so strong that transitions are rarely driven this way.

Transitions can also be driven by the time-variation of the light field. Let the molecule be at $z_1$ at time $t=t_1=\pi/\delta$, and linearize around this value of $t$. We obtain $\epsilon \approx \frac{1}{2}\hbar\zeta \delta \sin(2k z_1)(t-t_1)$ and $V \approx -\hbar \zeta \sin^2(k z_1).$ The probability of a diabatic transition at this crossing is
\begin{equation}
    P_{\rm LZ}^{(2)} \approx \exp\left(-2\pi \frac{\zeta}{|\delta|}\frac{\sin^4(k z_1)}{|\sin(2k z_1)|} \right).
    \label{eq:PLZ2}
\end{equation}
This probability is controlled by the dimensionless parameter $\zeta/\delta$ which is the ratio of the light shift to the rate of polarization modulation. When $\zeta/\delta \gg 1$, the diabatic transitions will be strongly localized around the points where $z_1 \approx m \pi/k$, while if $\zeta/\delta \ll 1$ there can be diabatic transitions almost anywhere (no stable dark state). Cooling will be most effective for intermediate values of $\zeta/\delta$. Our studies have focussed on the regime where $\zeta\sim \Gamma$ and $\delta\sim0.2 \Gamma$, so that $\zeta/\delta \sim 5$. This is the regime where transitions are localized around the bottom of the potential but not so strongly localized that they become improbable.

To summarize, we see that molecules tend to be optically pumped to the dark state near the tops of the potential hills, then make diabatic transitions back to the bright state near the bottom of the hills. These diabatic transitions can be driven by the motion when the velocity is not too small, and can be driven by the time variation of the field at any velocity. Molecules with larger velocities climb further up the hills, so lose more energy, resulting in a strong velocity-dependent force that damps their motion towards zero velocity.

\subsection{Cooling mechanism in a magnetic field \label{sec:lattice_trapping}}

A magnetic field lifts the degeneracy of the $m=\pm1$ states that form the dark state in Eq.~(\ref{eq:dark_superposition}), introducing an additional time-dependence into the dark state. In the Landau-Zener picture, the magnetic field shifts the space-time locations of the diabatic crossings away from the minima of the light shift, making the cooling towards zero velocity less effective. Once $|\omega_{\rm Z}|>\zeta/2$, $\epsilon$ no longer crosses zero and the cooling mechanism breaks down. Instead, there is a force that drives the molecules towards the velocity of one of the moving lattices. We examine the origin of this frictional force in the frame of the lattice. 

Consider a molecule moving through the counter-propagating lattices. It sees only one lattice at a time, depending on whether it has $m=\pm1$, and hops randomly between these lattices due to photon scattering. Switching state places the molecule at a random position in the new lattice, but (for $\Delta>0$) photon scattering always tends to happen at the top of the potential hill where the intensity is highest. Thus, in the frame of each lattice, a molecule tends to lose energy, resulting in a force that drives the velocity towards zero in the lattice frame. We will refer to the lattice moving in the same direction as the molecule as the co-lattice, and the one moving in the opposite direction as the counter-lattice. To generate any net force, we need to break the symmetry between the two lattices. There are two effects that do this, both arising from the motion of the molecule. The first effect is due to the different impulses exerted by the two lattices. Suppose the molecule has a speed close to $\delta/(2k)$. It moves slowly through the co-lattice so spends a long time climbing the potential hill, resulting in a large impulse that produces a large reduction in velocity (in the co-lattice frame). The molecule moves much more quickly through the counter-lattice resulting in a smaller impulse and a smaller reduction in velocity (in the counter-lattice frame). For a small change in kinetic energy $K$ the change in velocity is inversely proportional to the velocity, $\Delta v=K/(m v)$. This effect drags molecules towards the co-lattice. The second effect is due to the optical pumping probability. When the velocity is close to $\delta/2k$ molecules spend a long time at the top of the potential hills of the co-lattice, but ride quickly over the hills of the counter-lattice. The optical pumping probability is weighted towards the top of the hills in both lattices, since this is where the scattering rate is highest, but the weighting is stronger in the lattice where molecules move more slowly. Both effects drive molecules towards $\pm \delta/(2k)$, with the sign determined by the initial velocity.

\begin{figure}[tb]
    \includegraphics[width=0.9\linewidth]{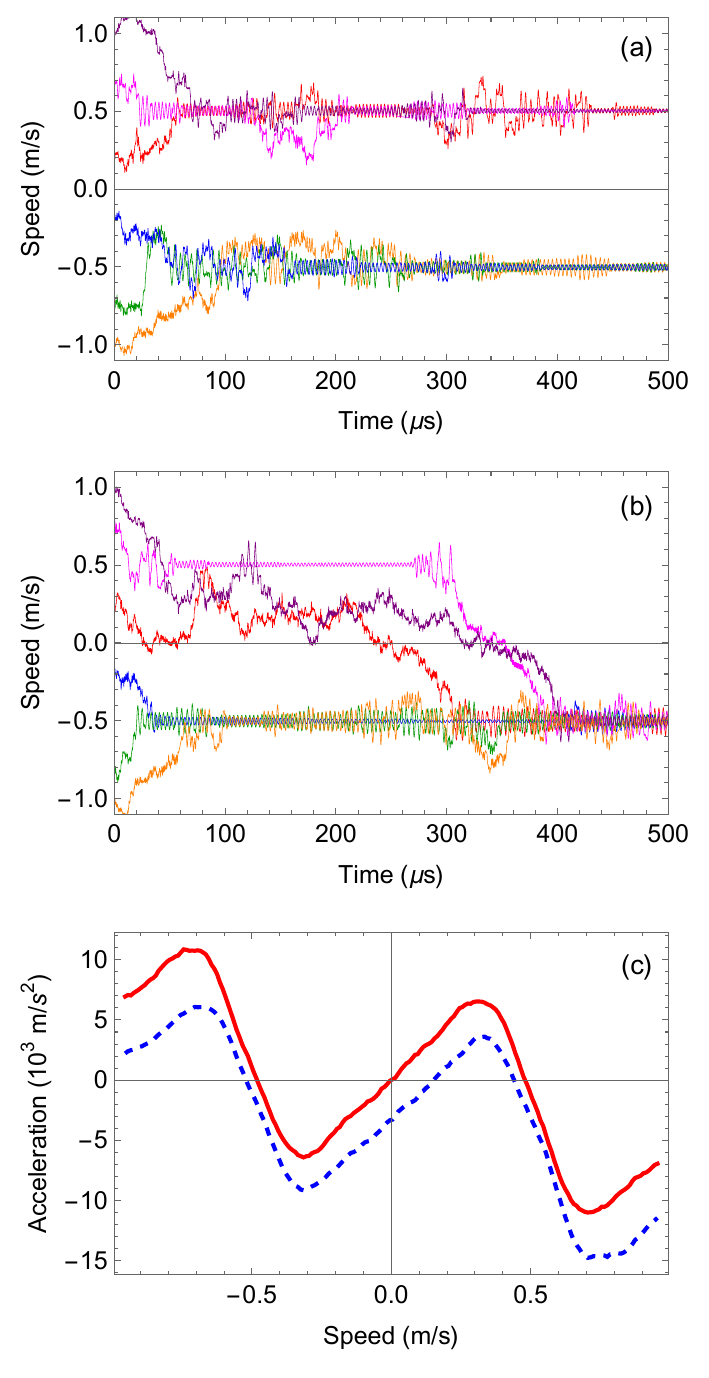}
    \caption{(a, b) Speed versus time for molecules with different initial speeds, found from the lattice hopping model described in section \ref{sec:lattice_trapping}. The parameters are $\lambda=606$~nm, $\Gamma=5.2\times 10^7$~rad/s, $M=59u$, $s_0=5$, $\Delta=2\Gamma$, $\delta=0.2\Gamma$ and (a) $F_0=0$, (b) $F_0=-6.7 \times 10^{-3} \hbar k \Gamma$. (c) Mean acceleration versus speed determined from this model for $F_0=0$ (solid red) and $F_0=-6.7 \times 10^{-3} \hbar k \Gamma$ (dashed blue).}
    \label{fig:trajectories}
\end{figure}

These processes are captured by a very simple 1D model of a molecule jumping at random between $m=\pm 1$ at a rate proportional to the scattering rate, $R=\frac{\Gamma}{2}\frac{s}{1+s+4\Delta^2/\Gamma^2}$. For each scattering event, the momentum changes by $\hbar k$ in a random direction. The molecule moves under the influence of the dipole force $F_{\rm dip}=-\frac{\hbar \Delta}{2}\frac{\nabla s}{1+s+4\Delta^2/\Gamma^2}$. For $m=\pm 1$, $s=s_0 \cos^2(k z \pm \delta t/2)$. We call this the {\it lattice hopping model}. Figure \ref{fig:trajectories}(a) shows some typical trajectories determined from this model in the case where $v_{\rm lattice}=\delta/(2k)\simeq 0.5$~m/s. The trajectories illustrate the effects described above, with molecules dragged to $\pm 0.5$~m/s with the sign determined by the initial velocity. Eventually, a molecule with speed very close to the lattice speed will, by chance, find itself in $m=\mp 1$ and at the node of the $\sigma^\pm$ standing wave, making only small oscillations around the node. In this situation, the molecule is hardly ever excited because it is dark to the $\sigma^\mp$ lattice and only ever sees low intensity $\sigma^\pm$ light. We see this in the trajectories which eventually settle to low amplitude oscillations around the lattice velocity.

The solid line in Fig.~\ref{fig:trajectories}(c) shows the mean acceleration determined from the lattice hopping model. For each speed, we simulate the motion for 10~$\mu$s, calculate the average acceleration over this interval, repeat 400 times, then average the results. We see that there is a force driving molecules away from zero velocity and damping the motion towards the lattice velocities. We also see that the force is stronger for $|v|>|v_{\rm lattice}|$ than for  $|v|<|v_{\rm lattice}|$. We attribute this to the combined effect of the co- and counter-lattices. The force from the counter-lattice reinforces the damping when the velocity is above that of the co-lattice, but weakens it below. The residual effect of the mechanism described in section \ref{sec:cooling_zero_B} also does this since it drives the motion towards zero velocity, though that is not captured by the lattice hopping model. 

So far, this model does not explain why one lattice is preferred over the other. It was previously suggested that this preference arises because the Zeeman shift introduces a difference in detuning that makes one lattice deeper than the other~\cite{Li2025}. Although the lattices do have different depths, this does not appear to play any role.  Instead, the role of the magnetic field is the ZIDS mechanism described in section \ref{sec:trapping_mechanism}. Until a molecule is trapped at the node of its co-moving lattice, this mechanism still applies a force. When $B>0$ and $\delta>0$, the force is negative at all velocities (see Fig.~\ref{fig:1D_force}). There is a range of magnetic fields where this force is large enough to drive molecules away from the lattice moving with positive velocity (which has weaker damping on the lower velocity side), but not large enough to drive them out of the lattice moving with negative velocity (which has stronger damping on the lower velocity side). Consequently, within this range of $|B|$, the velocity is driven towards $-\delta/(2k)$ when $B>0$ and towards $+\delta/(2k)$ when $B<0$. 

To capture this effect in the lattice hopping model, we simply add a constant force $F_0$ which we take as a free parameter. Figure \ref{fig:trajectories}(b) shows some trajectories determined in the case where $F_0=-6.7 \times 10^{-3} \hbar k \Gamma$. Molecules whose initial velocity is negative are rapidly driven to $-v_{\rm lattice}$ and then remain close to this speed as they oscillate around the nodes of the lattice, rarely switching state. Molecules whose initial velocity is positive may get trapped at $v_{\rm lattice}$ for a while, but this is unstable to fluctuations -- when the molecule comes out of the lattice it tends to be driven by $F_0$ towards negative velocities, and eventually ends up at $-v_{\rm lattice}$. The dashed line in Fig.~\ref{fig:trajectories}(c) shows the force in this case. We see that there is still damping towards $+v_{\rm lattice}$ but the range of speeds where this is stable is narrow. By contrast, molecules are damped to $-v_{\rm lattice}$ over a wide range of speeds. For the parameters used in Fig.~\ref{fig:trajectories} we find that trapping around $+v_{\rm lattice}$ is de-stabilized once $|F_0|\gtrsim 6.3 \times 10^{-3} \hbar k \Gamma$, while trapping around $-v_{\rm lattice}$ remains stable for $|F_0| \lesssim 13 \times 10^{-3} \hbar k \Gamma$.

\begin{figure}[tb]
    \includegraphics[width=0.9\linewidth]{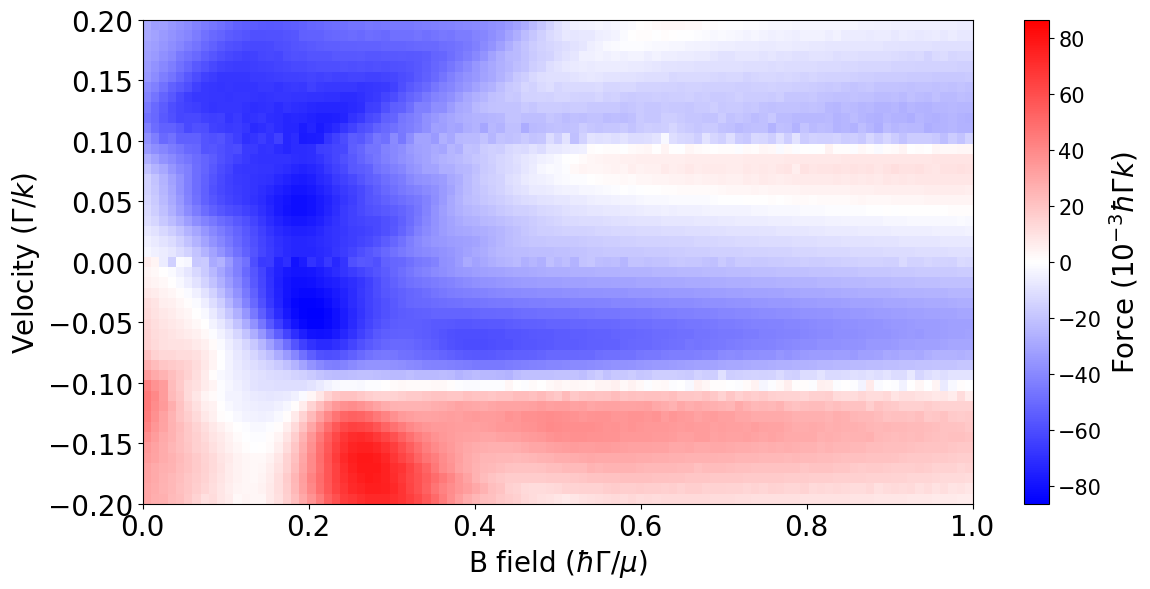}
    \caption{OBE simulation of the force map for $\lambda=606$~nm, $\Gamma=5.2\times 10^7$~rad/s, $M=59u$, $s_0=5$, $\Delta=2\Gamma$, $\delta=0.2\Gamma$.} 
    \label{fig:conv_heatmap}
\end{figure}

\begin{figure*}[t]
    \centering
    \subfloat[]{
        \includegraphics[width=0.25\textwidth]{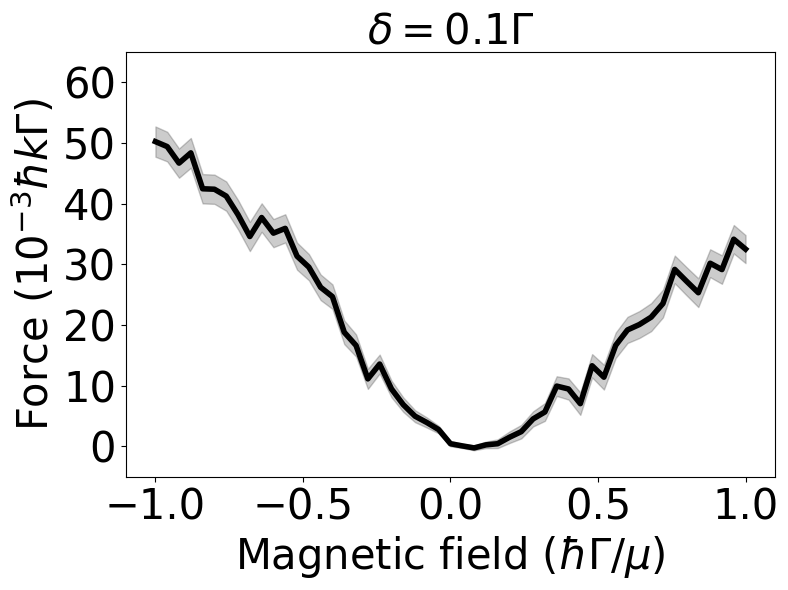}
        \label{fig:3D_single_dp1}
        }
    \subfloat[]{
        \includegraphics[width=0.25\textwidth]{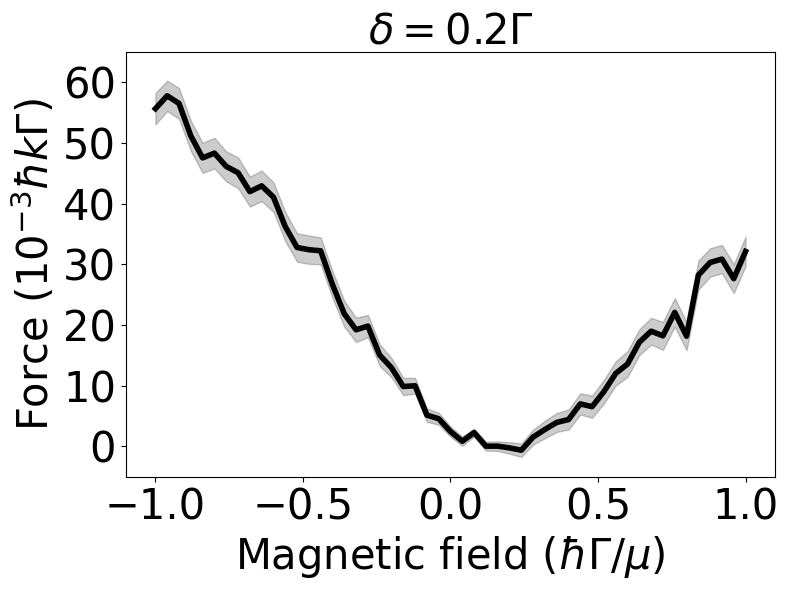}
        \label{fig:3D_single_dp2}
        }
    \subfloat[]{
        \includegraphics[width=0.25\textwidth]{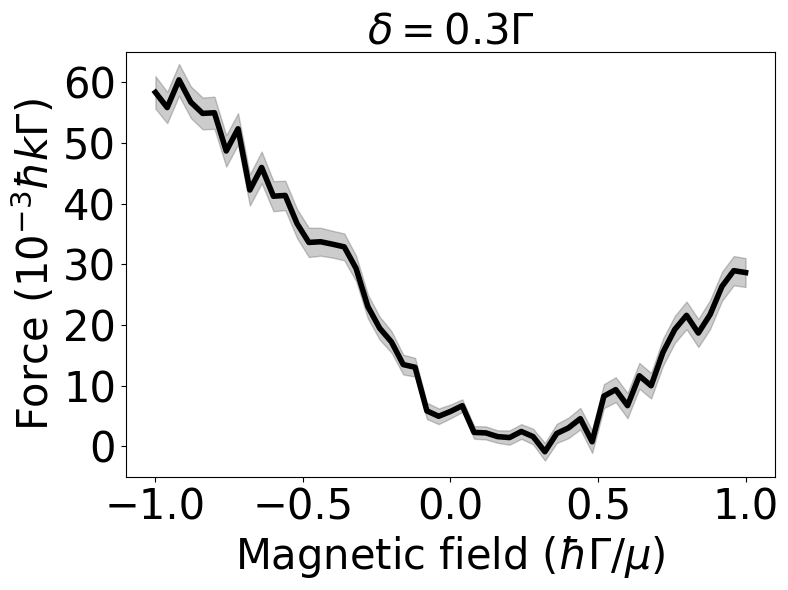}
        \label{fig:3D_single_dp3}
        }
    \hfill
    \subfloat[]{
        \includegraphics[width=0.25\textwidth]{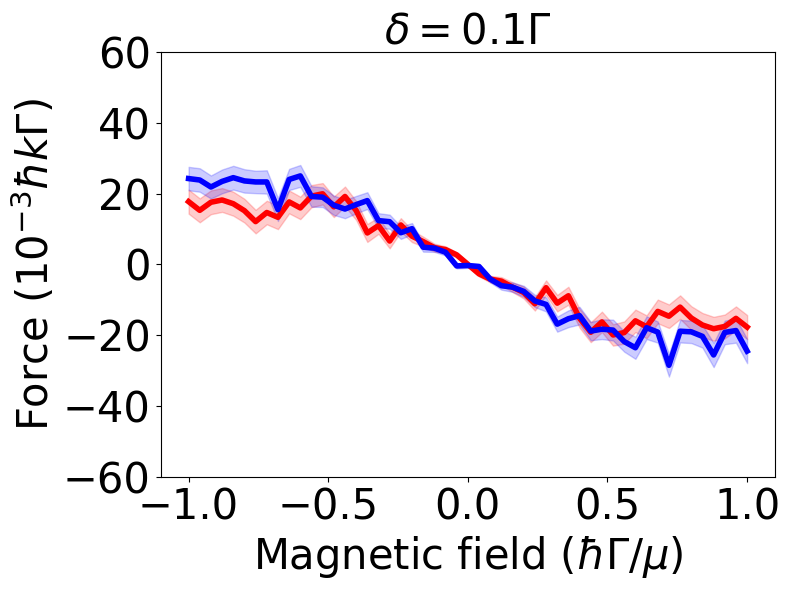}
        \label{fig:full_3D_dp1}
        }
   \subfloat[]{
        \includegraphics[width=0.25\textwidth]{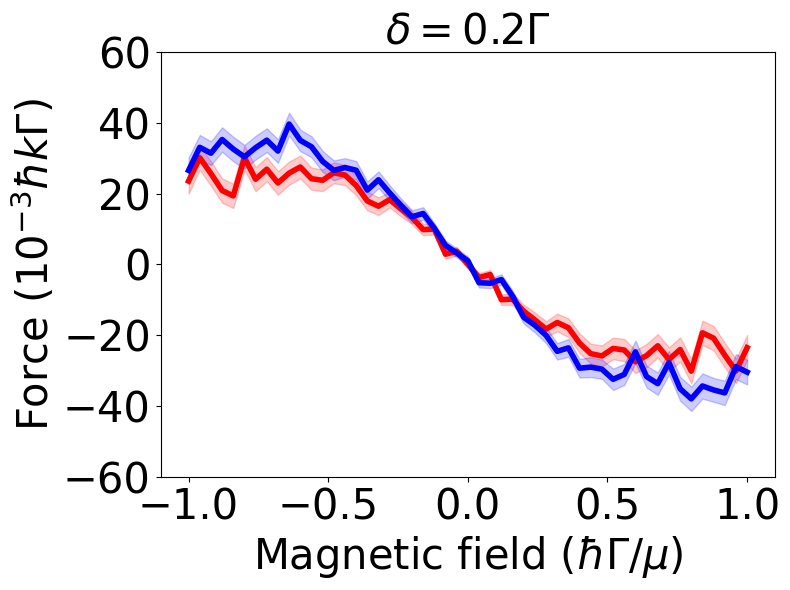}
        \label{fig:full_3D_dp2}
        }
   \subfloat[]{
        \includegraphics[width=0.25\textwidth]{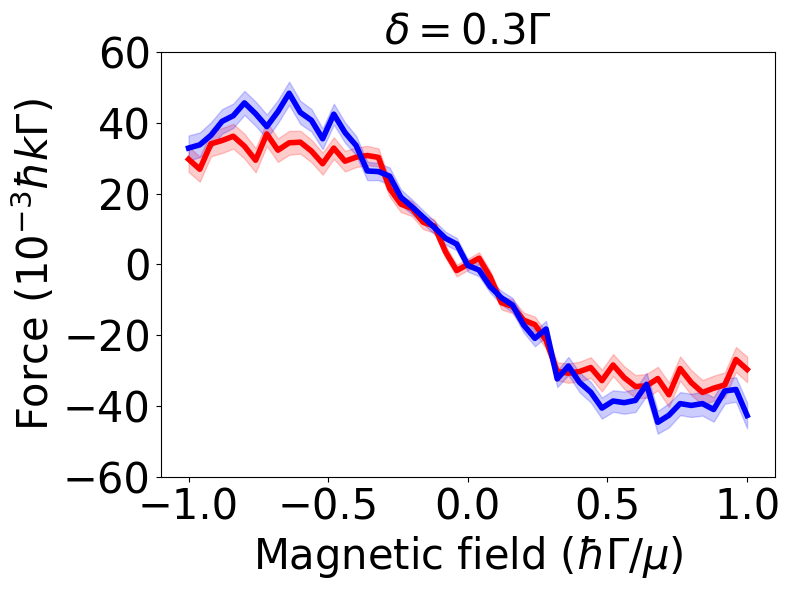}
        \label{fig:full_3D_dp3}
        }
    \caption{3D simulations of the force in the $z$-direction, as a function of magnetic field. The parameters are $\Delta=2\Gamma$, $s=5$ per beam and $N_{\rm rep}=1024$. (a-c) $F_{z+}$ for three values of $\delta$. (d-f) Force with all six beams present (blue) compared to $F_{z+}+F_{z-}$ (red).}
    \label{fig:full_3D}
\end{figure*}

Figure \ref{fig:conv_heatmap} shows the result of the OBE simulation for the same parameters as used for Fig.~\ref{fig:trajectories}. The simulation extends to higher values of $B$ compared to those in Fig.~\ref{fig:1D_heatmap}. We see that at low $B$ the molecules are driven towards $-v_{\rm lattice}$ (similar to Fig.~\ref{fig:trajectories}(b)), but as $B$ increases the trapping force weakens and once $\mu B \gtrsim 0.5 \hbar\Gamma$ the molecules can be confined in either of the two lattices (similar to Fig.~\ref{fig:trajectories}(a)). The range of forces where only one lattice is stable is similar to the range found above. Thus, we see that our very simple lattice hopping model captures all the essential physics of the `conveyor belt' mechanism. 

The results imply that molecules can easily be lost from the blue-detuned MOT. Those molecules with $v$ close to $+v_{\rm lattice}$ and at large enough $B$ will be confined at $+v_{\rm lattice}$ and driven out of the MOT. This is a relevant process in existing blue-detuned molecule MOTs. For example, when $\mu=\mu_{\rm B}$, $\Gamma=5.2\times 10^7$~rad/s, and $dB/dz=30$~G/cm, a molecule at $z=1$~mm will have $\mu B = 0.5 \hbar\Gamma$. This suggests that, to avoid losses, the magnetic field gradient needs to be ramped up slowly so that the MOT is compressed without too many molecules seeing large fields. The timescale is determined by the time needed to reach equilibrium. The blue-detuned MOT behaves as a heavily overdamped oscillator, so the relevant timescale is $\gamma/\omega_0^2$, where $\gamma$ is the damping constant and $\omega_0$ is the oscillation frequency. For the parameters explored in the next section, this characteristic timescale is about 1~ms.

\section{Model in three dimensions}\label{sec:model_3D}
\subsection{Trapping mechanism}

Next, we study the $F=1$ - $F'=1$ system in three dimensions by adding beams propagating in the $x$ and $y$ directions, identical to those along $z$. To see whether the ZIDS trapping mechanism persists, we first remove the beams propagating towards $-z$ and calculate the force in the $z$-direction, $F_{z+}$,at $v=0$ and as a function of $B$. Figures \ref{fig:full_3D}(a-c) show the results for three different values of $\delta$, where we have fixed $\Delta=2\Gamma$ and $s=5$ per frequency per beam. We see that, just as in 1D, $F_{z+}$ goes to zero close to the critical magnetic field where $\mu B=\hbar \delta/2$, verifying that the mechanism is still effective in 3D. The feature is substantially broader in 3D than the one found in 1D. Similarly, we calculate $F_{z-}$ by removing the beam propagating towards $+z$. Figures \ref{fig:full_3D}(d-f) compares the force calculated for all six beams, to $F_{z+}+F_{z-}$. We see very good agreement between these approaches, indicating that the force in 3D can be understood as arising almost entirely from Zeeman-induced dark states generated by pairs of co-propagating beams. The moving lattices produced by the counter-propagating beams do not contribute to the trapping.

\subsection{Spring and damping constants}
\label{subsection:1to1_characterization}

\begin{figure*}
    \subfloat[]{
        \includegraphics[width=0.3\linewidth]{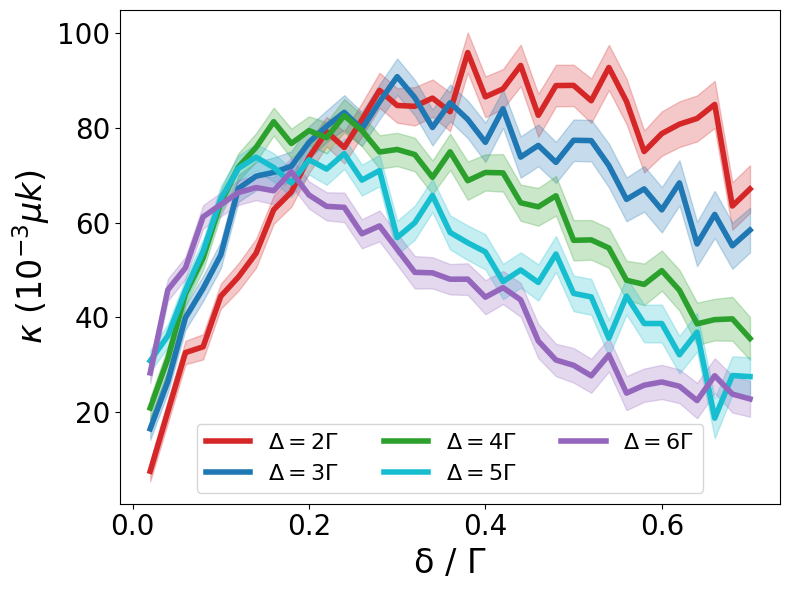}
        \label{fig:spring_var_delta}
        }
    \subfloat[]{
        \includegraphics[width=0.3\linewidth]{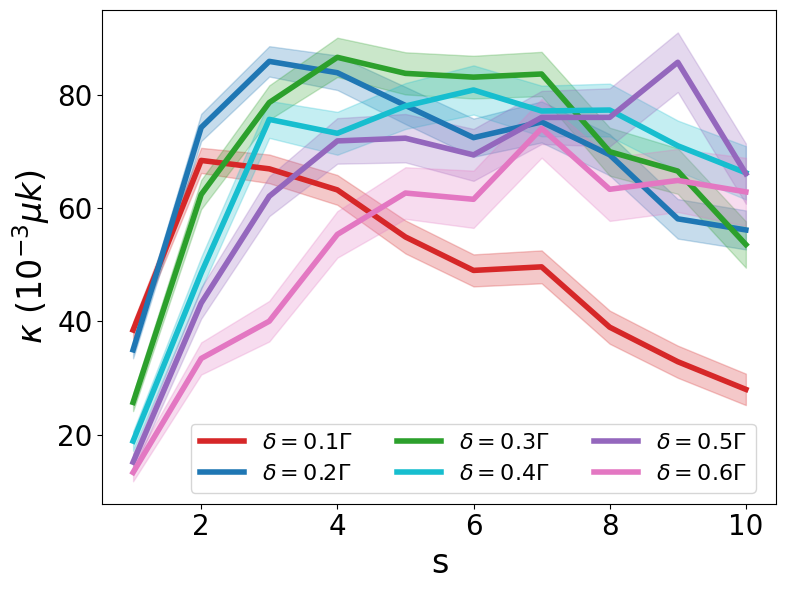}
        \label{fig:spring_var_s}
        }
    \subfloat[]{
        \includegraphics[width=0.3\linewidth]{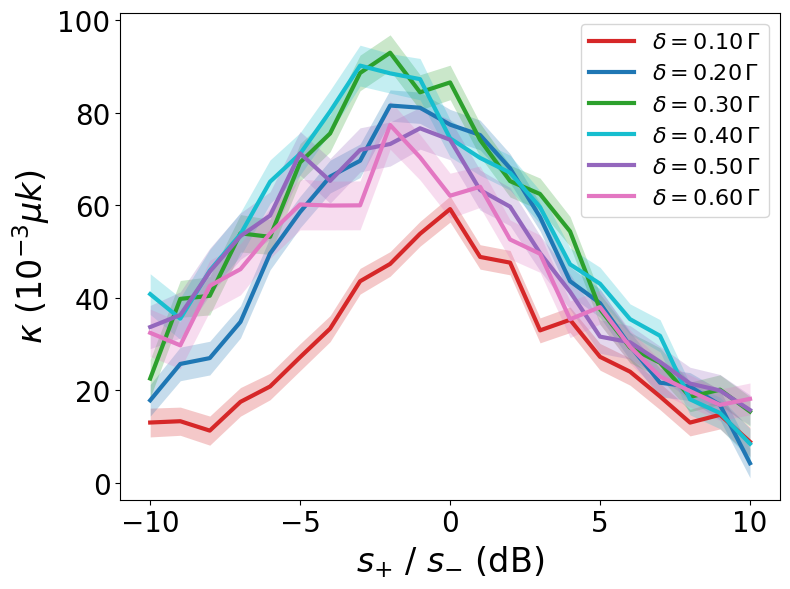}
        \label{fig:spring_var_sratio}
        }
    \hfill
    \subfloat[]{
        \includegraphics[width=0.3\linewidth]{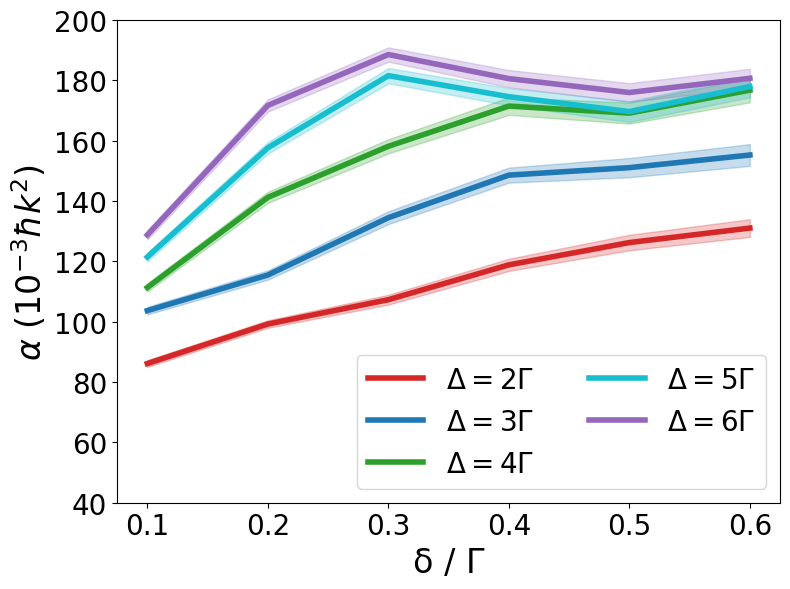}
        \label{fig:damping_var_delta}
        }
    \subfloat[]{
        \includegraphics[width=0.3\linewidth]{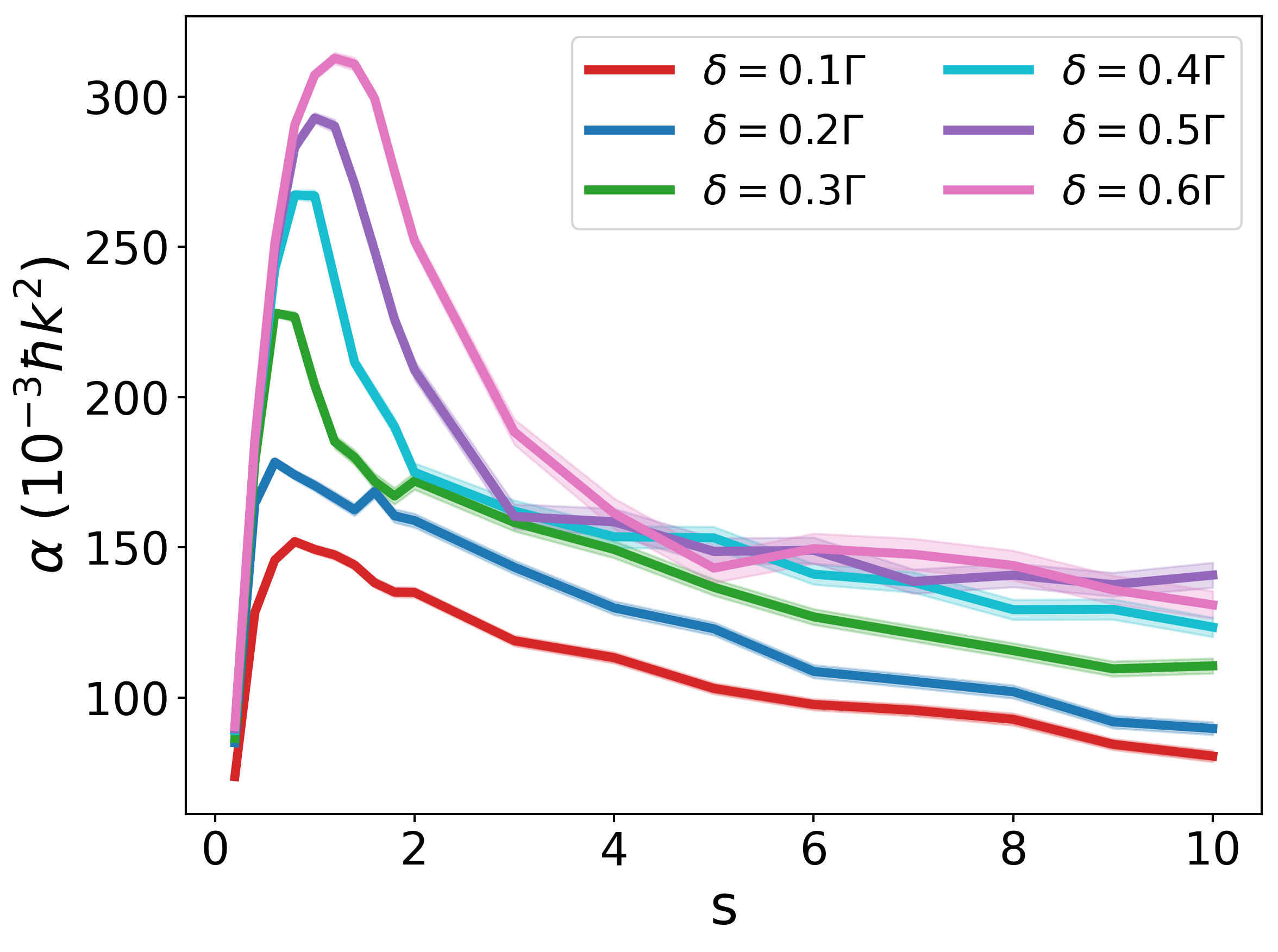}
        \label{fig:damping_var_s}
        }
    \subfloat[]{
        \includegraphics[width=0.3\linewidth]{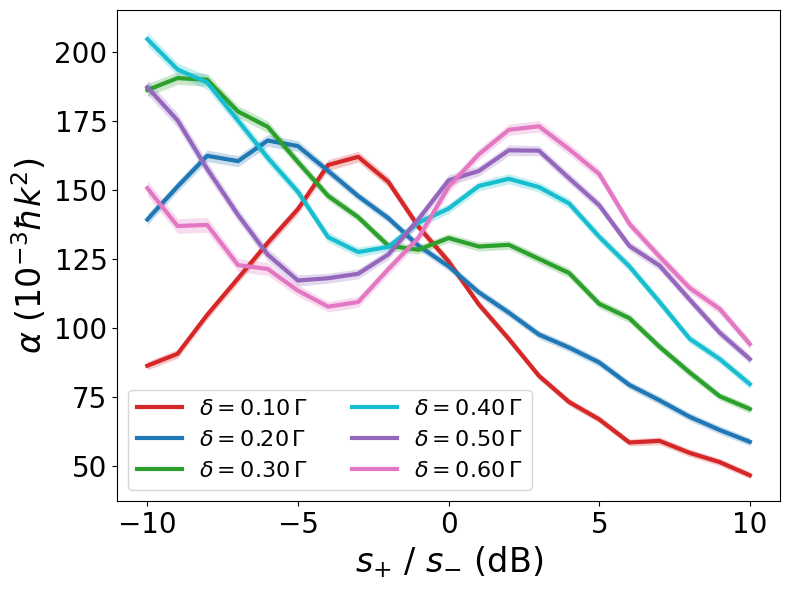}
        \label{fig:damping_var_sratio}
        }
     \caption{Characterization of the 3D blue-detuned MOT for the $F=1$ - $F'=1$ system, showing the spring constant, $\kappa=(\partial F/\partial B)_{v=0}$ and damping constant $\alpha=(\partial F/\partial v)_{B=0}$. (a, d) $\kappa$ and $\alpha$ versus $\delta$ for several values of $\Delta$. We have fixed $s=5$ per frequency per beam. (b, e) $\kappa$ and $\alpha$ versus $s$ for several values of $\delta$, with $\Delta=3\Gamma$. (c, f) $\kappa$ and $\alpha$ versus the intensity ratio $s_{+}/s_{-}$, with $\Delta=3\Gamma$ and a total $s=60$ summed over all frequency components and all six beams.
    }
    \label{fig:characterization}
\end{figure*}

The restoring force in a MOT is characterized by the spring constant, $\kappa$, and the cooling can be characterized by the damping constant, $\alpha$. From the simulations, we calculate $\partial F/\partial B$ at zero velocity and refer to this as the `spring constant' since $\partial F/\partial z$ follows from this quantity once the magnetic field gradient is specified. Similarly, the damping constant is determined from $\partial F/\partial v$ at zero magnetic field.

Figure \ref{fig:characterization}(a) shows how $\kappa$ varies with $\delta$, for several different choices of $\Delta$. We see that $\kappa$ rises rapidly as $\delta$ increases from zero, then reaches a maximum and declines more slowly as $\delta$ increases further. Increasing $\Delta$ lowers the maximum value of $\kappa$ (though only a little), shifts the maximum to lower values of $\delta$, and yields a narrower peak. Figure \ref{fig:characterization}(b) shows that $\kappa$ is relatively insensitive to $s$ once $s$ exceeds a threshold value. This threshold increases with increasing $\delta$. Figure \ref{fig:characterization}(c) plots $\kappa$ as a function of the intensity ratio of the two frequency components $s_{+}/s_{-}$. We see that there is a small asymmetry, with the highest spring constant typically found where $s_{+} \approx s_{-}/2$. 

Figure \ref{fig:characterization}(d) shows how $\alpha$ depends on $\delta$ and $\Delta$, when the intensity in each component is $s=5$. To understand the behaviour observed, note that the damping constant is determined from the gradient of the force near zero velocity, so that diabatic transitions are mostly driven by the time-variation of the light field rather than by motion through the light field, and the most relevant parameter is the ratio $\zeta/\delta \propto s\Gamma^2/(\delta \Delta)$. We see that $\alpha$ first increases with $\delta$ and then reaches a plateau. At larger values of $\Delta$ the plateau is reached at smaller $\delta$, consistent with the idea that there is an optimum value for the parameter $s\Gamma^2/(\delta \Delta)$.  Figure \ref{fig:characterization}(e) shows that $\alpha$ has a pronounced peak near $s\approx1$. As $\delta$ increases, this peak gets larger and shifts to higher $s$. With the exception of $\delta=0.1\Gamma$, the peak is located at $s\approx 2.2 \delta/\Gamma$. This is again consistent with there being an optimum for $s\Gamma^2/(\delta \Delta)$. Increasing $s$ increases the height of the potential hills, so as the optimum $s$ increases the maximum value of $\alpha$ also increases. Beyond $s>3$ there is a much weaker dependence of $\alpha$ on $s$. Figure \ref{fig:characterization}(f) shows that $\alpha$ has a surprising dependence on the intensity ratio $s_{+}/s_{-}$. At low $\delta$ there is a peak at $s_{+}/s_{-}<1$, with the peak shifting to smaller intensity ratios as $\delta$ increases. At larger $\delta$ however, a new peak appears where $s_{+}\approx 2 s_{-}$. This becomes more pronounced as $\delta$ increases, but remains at roughly the same intensity ratio. Taking Figs.~\ref{fig:characterization}(c) and (f) together, we see that there is a range of intensity ratios around $s_{+} \approx s_{-}$ where the spring constant and damping constant are both large.

\section{Model with two ground hyperfine states}\label{sec:model_two_hyperfine}

\begin{figure}[tb]
    \subfloat[]{
        \includegraphics[width=0.28\linewidth]{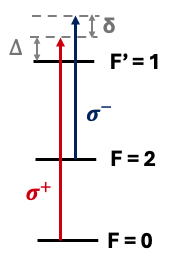}
        \label{fig:lambda_mot_scheme}
        }
    \subfloat[]{
        \includegraphics[width=0.65\linewidth]{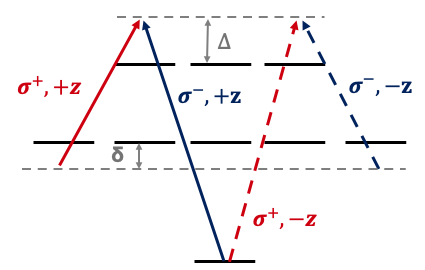}
        \label{fig:lambda_laser_config}
        }
    \hfill
    \subfloat[]{
        \includegraphics[width=0.48\linewidth]{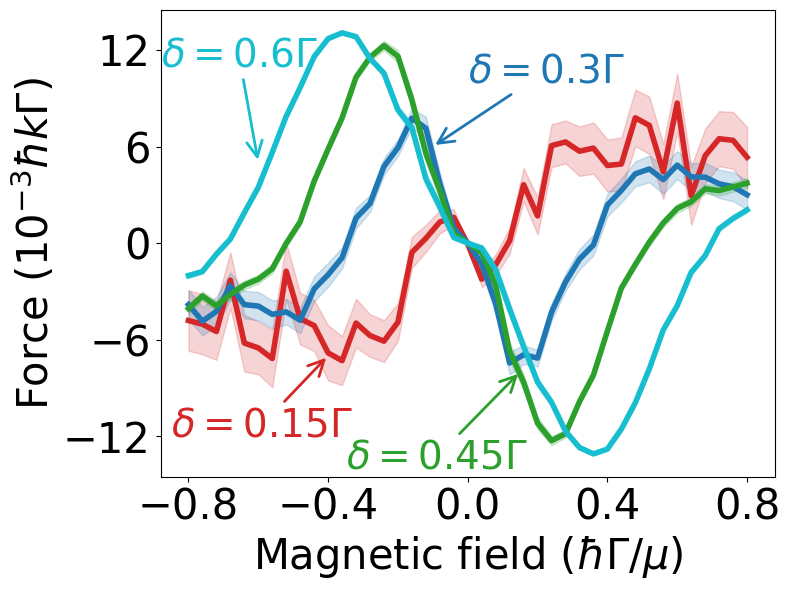}
        \label{fig:lambda_trapping_var_B}
        }
    \subfloat[]{
        \includegraphics[width=0.48\linewidth]{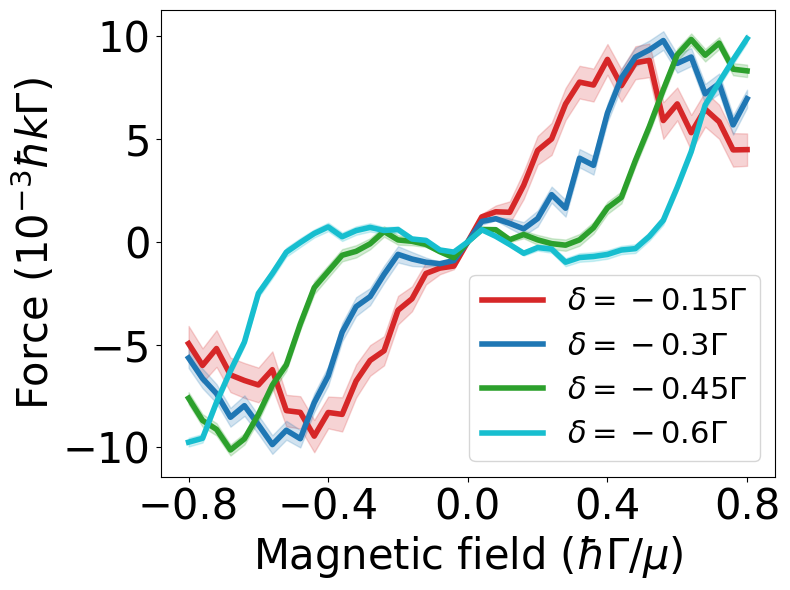}
        \label{fig:lambda_anit_trapping}
        }
    \caption{(a) Blue-detuned MOT scheme with two well-separated ground-state hyperfine levels, $F=0$ and 2, and a single excited state $F'=1$. $\Delta$ is the global detuning and $\delta$ is the two photon detuning. The figure shows the polarizations of the lasers propagating towards $-z$, and we take $dB/dz>0$. (b) Illustration of Zeeman-induced dark state formed by the pair of beams travelling towards $+z$ (solid arrows), and towards $-z$ (dashed arrows) (c) Force versus magnetic field for $\Delta=2\Gamma$, $s=5$ per frequency per beam in 1D, and various positive values of $\delta$. (d) Same as (c) but for negative $\delta$.}
    \label{}
\end{figure}

The mechanisms described so far can also work between two different hyperfine components of the ground state. We consider a model with $\ket{F=2}$ and $\ket{F=0}$ in the ground state and $\ket{F'=1}$ in the excited state. As before, the magnetic moment is $\mu$ in the ground state, defined so that the Zeeman shift of the $\ket{2,2}$ state is $\mu B$. The excited state decays with equal probability to $F=0$ and $F=2$, and the total decay rate is $\Gamma$. 

\subsection{Trapping mechanism}
We start from the 1D configuration shown in Fig.~\ref{fig:lambda_mot_scheme}. The two frequency components have opposite circular polarizations, a global detuning $\Delta$ and a frequency difference of $\delta$. First consider the case where only the laser propagating towards $+z$ is present. The states $\ket{2,+1}$ and $\ket{2,+2}$ cannot couple to the light at any $B$. In addition, as can be seen from the solid arrows in Fig.~\ref{fig:lambda_laser_config}, there is a a Zeeman-induced dark state $\ket{d_+}=\alpha e^{i(\omega_{\rm Z} - \delta)t}\ket{0,0}+\beta \ket{2,-2}$, where $\alpha, \beta$ depend on the Rabi frequencies. At the critical magnetic field $B_{\rm c}=\hbar\delta/(2\mu)$, $\ket{d_+}$ becomes time independent and the molecule cannot couple to the $+z$ laser beam, but will still couple to the $-z$ beam. The resulting imbalance in radiation pressure produces a $B$-dependent force, exactly as for the 1-1 system studied earlier. This mechanism is likely to be important in several of the blue-detuned molecule MOTs reported previously~\cite{Burau2023, Li2024, Jorapur2024, Yu2026, Hallas2024arxiv, Zeng2025arxiv}.

Figure \ref{fig:lambda_trapping_var_B} shows the force as a function of the magnetic field in this 1D system, for various positive values of $\delta$. As expected, the force has turning points near $\pm B_{\rm c}$. The maximum force increases and shifts to higher $|B|$ with increasing $\delta$. This is similar to the behaviour of the 1-1 system studied earlier. Figure \ref{fig:lambda_anit_trapping} shows the force for negative values of $\delta$. While the behaviour is similar, there is some asymmetry between the $\delta>0$ and $\delta<0$ cases. For $\delta<0$ the force tends to be smaller and the turning points are shifted to higher values of $|B|$. We suppose this asymmetry arises because $\delta$ is comparable to $\Delta$, so makes an appreciable difference to the overall detuning of the light.

\begin{figure*}[tb]
    \subfloat[]{
        \includegraphics[width=0.3\linewidth]{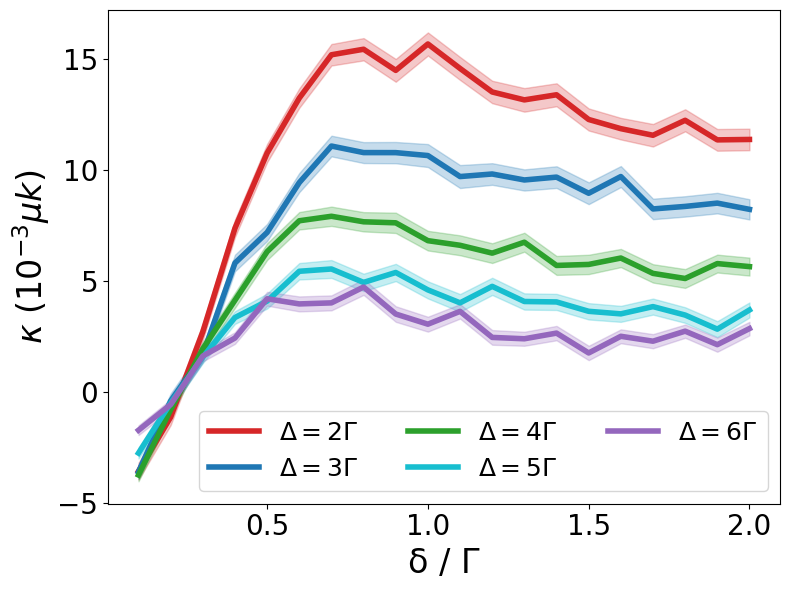}
        \label{fig:lambda_spring_var_delta}
        }
    \subfloat[]{
        \includegraphics[width=0.3\linewidth]{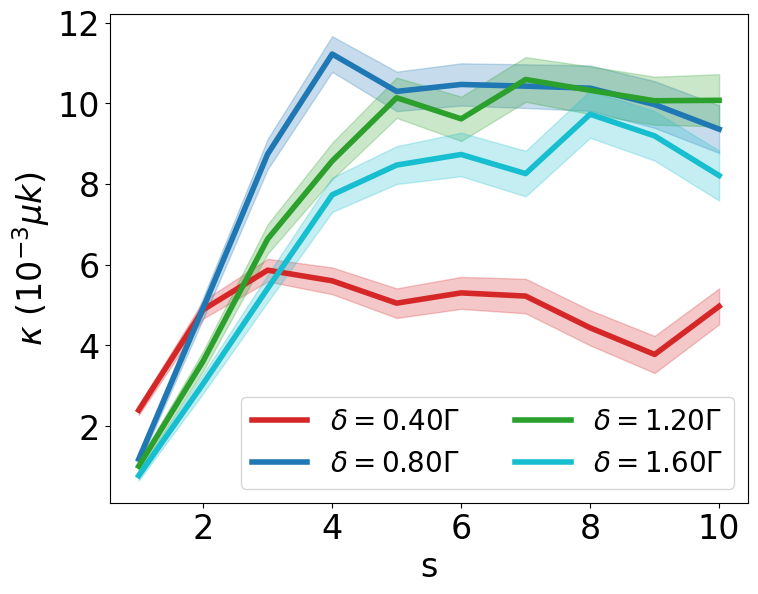}
        \label{fig:lambda_spring_var_s}
        }
    \subfloat[]{
        \includegraphics[width=0.3\linewidth]{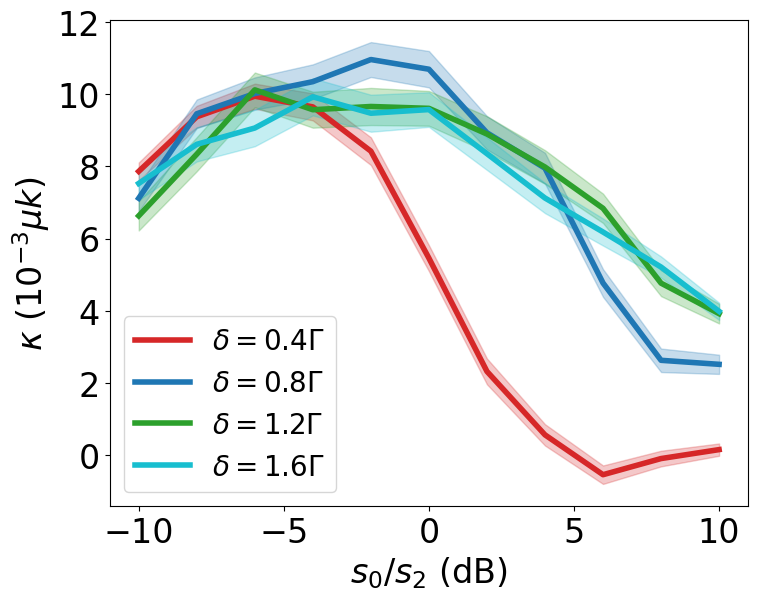}
        \label{fig:lambda_spring_var_sratio}
        }
    \hfill
    \subfloat[]{
        \includegraphics[width=0.3\linewidth]{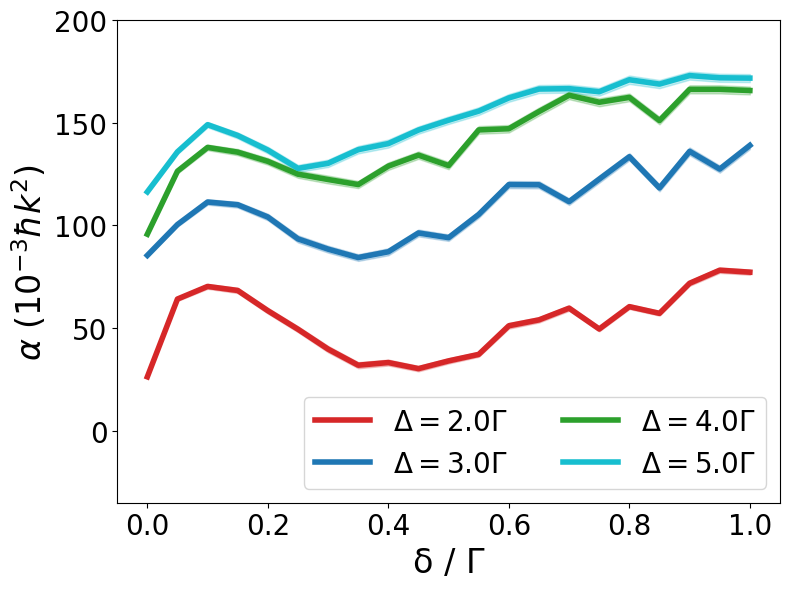}
        \label{fig:lambda_damping_var_delta}
        }
    \subfloat[]{
        \includegraphics[width=0.3\linewidth]{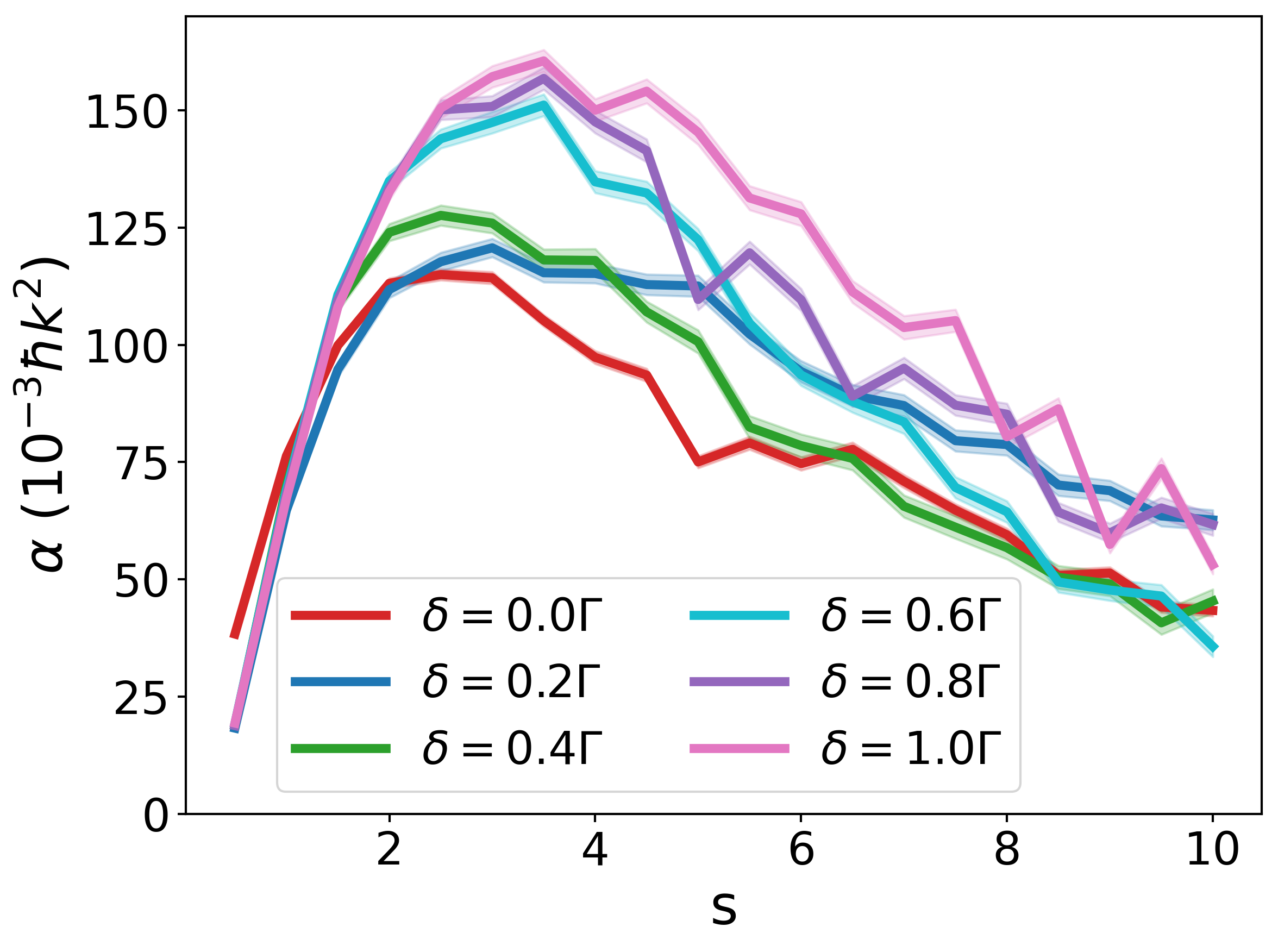}
        \label{fig:lambda_damping_var_s}
        }
    \subfloat[]{
        \includegraphics[width=0.3\linewidth]{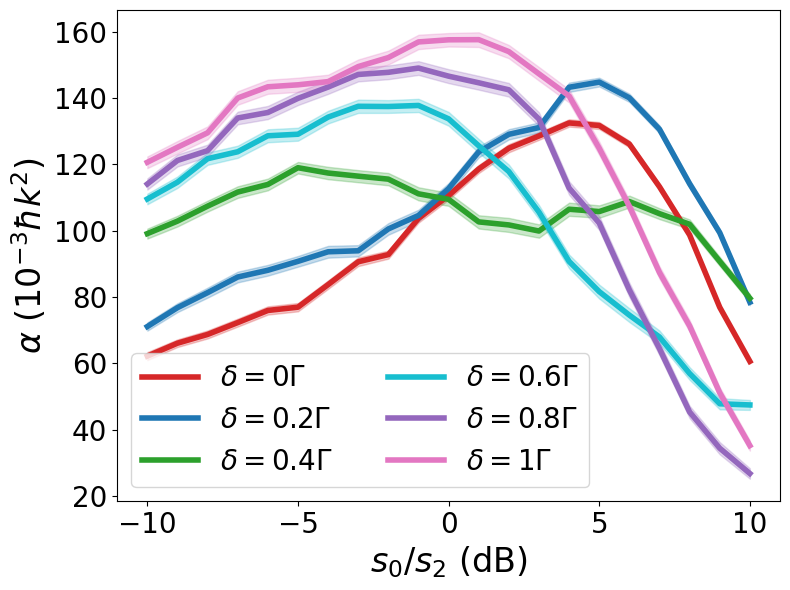}
        \label{fig:lambda_damping_var_sratio}
        }
    \caption{Characterization of a 3D blue-detuned MOT with configuration shown in~Fig.~\ref{fig:lambda_mot_scheme}. (a) Spring constant, $\kappa$, and (d) damping constant, $\alpha$, as a function of $\delta$ for various $\Delta$. The intensity is $s=5$ per frequency per beam. (b) $\kappa$ and (e) $\alpha$ as a function of $s$, for various $\delta$ and with $\Delta=3\Gamma$. (c) $\kappa$ and (f) $\alpha$ as a function of the intensity ratio $s_{0}/s_{2}$, where $s_{F}$ is the saturation parameter of the laser addressing hyperfine state $F$. Fixed parameters are $\Delta=3\Gamma$ and a total intensity $s_{\rm tot}=60$ for all frequency components and all three directions.}
    \label{fig:lambda_021_characterization}
\end{figure*}

\subsection{Spring and damping constants}

Once again, we characterize the MOT by determining the spring constant $\kappa$ and damping constant $\alpha$. Figure \ref{fig:lambda_spring_var_delta} shows that $\kappa$ rises with $\delta$ and plateaus at $\delta \approx 0.75\Gamma$. $\kappa$ is higher for lower $\Delta$ for all values of $\delta$. Figure \ref{fig:lambda_damping_var_delta} shows that $\alpha$ is relatively insensitive to $\delta$, and $\alpha$ increases as $\Delta$ increases, and reaches maximum values at $\Delta\sim4\Gamma$. Figures \ref{fig:lambda_spring_var_s} and \ref{fig:lambda_damping_var_s} show how $\kappa$ and $\alpha$ depend on $s$ for various $\delta$, with $\Delta=3\Gamma$. $\kappa$ initially rises rapidly with $s$ before reaching a maximum around $s\approx 4$ and then saturates, while $\alpha$ reaches a peak at $s\approx3$ then slowly goes down. Figure \ref{fig:lambda_spring_var_sratio} shows how the spring constant depends on the intensity ratio $s_{0}/s_{2}$, where $s_{F}$ is the saturation parameter of the laser addressing hyperfine state $F$. We see that a higher intensity for the $\ket{F=2}\rightarrow\ket{F'=1}$ transition is preferred for trapping. We attribute the behaviour seen in this plot to an effective change of $\delta$ arising from the differential ac Stark shift, so that it echoes Fig.~\ref{fig:lambda_spring_var_delta}. Increasing $s_{F=2}$ increases the differential Stark shift which increases the effective value of $\delta$. Consider first the region of the plot where $s_{F=2}>s_{F=0}$. Here, increasing $s_{F=2}$ takes $\delta$ further into the plateau region of Fig.~\ref{fig:lambda_spring_var_delta}, so $\kappa$ is relatively insensitive to the intensity ratio. Now consider the region where $s_{F=0}>s_{F=2}$. Here, increasing $s_{F=0}$ reduces the effective $\delta$, bringing it into the small-$\delta$ region where $\kappa$ decreases. Figure \ref{fig:lambda_damping_var_sratio} shows that $\alpha$ has a similar response to the intensity ratio and is maximized close to $s_0=s_2$.

\section{Conclusion}

We have studied blue-detuned MOTs driven by a pair of closely-spaced frequency components of opposite polarization where the light field can be viewed as a pair of oppositely polarized lattices propagating with speeds $\pm \delta/(2k)$. We have identified the mechanisms responsible for trapping and cooling in these MOTs.

The trapping is due to the Zeeman-induced dark state (ZIDS) mechanism. When the Zeeman splitting matches the detuning between the frequency components, there is a state that is dark to the beam propagating in one direction, but not to the beam from the opposite direction. This sets up an imbalanced radiation pressure which confines molecules around the magnetic field zero. 

At $B=0$ molecules are cooled towards zero velocity by gray molasses cooling. Notably, the time-varying polarization of the light field drives non-adiabatic transitions from dark to bright states, unlike standard gray molasses where these transitions are driven only by motion through the light field. When $|B|>B_{\rm hi}$ there is a velocity dependent force that drives molecules to the lattice speeds $\pm \delta/(2k)$, with the sign determined by the sign of the initial velocity. In this regime, some molecules will be transported towards the MOT centre, and some away from the centre, potentially resulting in loss from the MOT.  There is a range of magnetic fields $B_{\rm lo}<|B|<B_{\rm hi}$ where the velocity-independent force from the ZIDS mechanism adds to the velocity-dependent force from the lattices in such a way that molecules are always driven to one of the two lattices. In this regime, with the correct sign of field gradient, molecules will always be transported towards the centre of the MOT. Finally, when $0<|B|<B_{\rm lo}$, there is a competition between the force towards $v=0$ and the one towards $v=\delta/(2k)$, resulting in a zero-crossing of the force at an intermediate $v$. For the parameters studied in this paper, we find $B_{\rm lo} \approx 0.2 \hbar\Gamma/\mu$ and $B_{\rm hi} \approx 0.5 \hbar\Gamma/\mu$.

As noted already, there is a loss mechanism in the blue-detuned MOT because at high magnetic fields molecules can be transported out of the MOT by the moving lattice. This may be partly responsible for the short lifetimes observed at high field gradients in some blue-detuned MOTs, e.g. \cite{Li2024}. This observation suggests that, when transferring molecules into the blue-detuned MOT, it may be beneficial to ramp up the magnetic field slowly so that molecules remain always in the low magnetic field regime. The 3D blue-detuned MOT analyzed in section \ref{sec:model_two_hyperfine} has a maximum value of $\kappa$ of about $10^{-2}\mu k$. For a CaF molecule in a field gradient of 15~G/cm, and taking an average $\mu \approx 0.5 \mu_{B}$, this corresponds to a trap oscillation frequency of about 860~rad/s. This is similar to the trap frequencies measured in a red-detuned MOT of CaF~\cite{Williams2017}, and not so different from the values measured in a blue-detuned MOT of CaF (about 450~rad/s at this field gradient, see figure 3(d) of reference \cite{Li2024}). The typical damping constant found in section \ref{sec:model_two_hyperfine} is around $10^{-1}\hbar k^2$, which for CaF gives $\alpha/m \approx 1.2\times 10^4$~s$^{-1}$. This is about 20 times higher than measured in red-detuned MOTs~\cite{Williams2017} but only 3 times higher than measured for a blue-detuned MOT of CaF (see figure 4(a) of reference \cite{Li2024}). The larger damping compared to a red-detuned MOT is to be expected since the blue-detuned MOT relies on sub-Doppler processes. The large damping and reduced photon scattering associated with the dark states is responsible for the low temperatures of the blue-detuned MOT. The low temperature and high spring constant are responsible for the small size of the MOT, yielding the desired high density samples. 

\begin{acknowledgements}

We thank Lajos Palanki and Dylan Brown for their help with the numerical simulations. We also thank Tingkun Chen for early contributions to this project and Chris Ho, Stefan Truppe, Ben Sauer and Christian Hallas for inspiring discussions on the trapping and cooling mechanisms. This work has been supported by EPSRC through grants EP/W00299X/1, EP/V011499/1, EP/Z535898/1 and UKRI2226.

\end{acknowledgements}
\clearpage

\bibliographystyle{apsrev4-2}
\bibliography{references}

\end{document}